\documentclass[11pt]{article}
\usepackage[textwidth=450pt,textheight=650pt]{geometry}
\usepackage{amsmath}
\usepackage{amssymb}
\usepackage{bbold}
\usepackage{color}
\usepackage{titlesec}
\usepackage[bottom]{footmisc}
\usepackage[numbers,sort&compress]{natbib}
\usepackage{hyperref}
\usepackage{hypernat}
\usepackage{graphicx}
\usepackage{xspace}
\usepackage{tabularx}
\usepackage{relsize}
\usepackage{caption}

\DeclareMathOperator{\Tr}{Tr}
\newcommand\figref[1]{figure~\ref{#1}}

\makeatletter
\renewcommand\theequation{\thesection.\arabic{equation}}
\@addtoreset{equation}{section}
\makeatother
\setcaptionwidth{.9\textwidth}
\titleformat{\section}[block]
  {\large\bfseries}{\thesection.}{.3em}{}
  
\begin{document}
\pagestyle{empty}
\begin{flushright}
DCPT-08/33, ITP-UU-08/31, SPIN-08/23\\
June 5th, 2008
\end{flushright}
\vskip 7ex

\begin{center}
\begin{minipage}{.95\textwidth}
{\huge\bf Group theory of icosahedral virus capsids:\\[1ex] a
  dynamical top-down approach}\\[5ex] 
{\large\bf Kasper Peeters$^{1}$ and Anne Taormina$^{2}$}\\[5ex]
\llap{$^1$~}Institute for Theoretical Physics\\
Utrecht University\\
P.O.~Box 80.195\\
3508 TD Utrecht\\
The Netherlands\\[3ex]
\llap{$^2$~}Department for Mathematical Sciences\\
Durham University\\
South Road\\
Durham DH1 3LE\\
United Kingdom\\[3ex]
{\tt kasper.peeters@aei.mpg.de}, {\tt anne.taormina@durham.ac.uk}
\vskip 9ex

{\bf Abstract:}\\[1ex]
We explore the use of a top-down approach to analyse the dynamics of
icosahedral virus capsids and complement the information obtained from
bottom-up studies of viral vibrations available in the literature. A
normal mode analysis based on protein association energies is used to
study the frequency spectrum, in which we reveal a universal plateau
of low-frequency modes shared by a large class of Caspar-Klug
capsids. These modes break icosahedral symmetry and are potentially
relevant to the genome release mechanism. We comment on the role of
viral tiling theory in such dynamical considerations.
 

\end{minipage}
\vfill
\end{center}
\eject

\pagestyle{plain}
\addtocounter{page}{-1}
\hrule
\tableofcontents
\vspace{4ex}
\hrule

\vspace{3ex}

\section{Introduction and summary}

Viral protein capsids exhibit a surprising amount of symmetry, the
consequences of which are still far from fully explored. A major step
forward in the understanding of the \emph{static} properties of
icosahedral virus capsids was the introduction of viral tiling
theory~\cite{Twarock:2004a}. Viral tiling makes use of
non-crystallographic Coxeter groups \cite{Humphreys:1990} to predict
the distribution of proteins on the faces of the capsid. It fills an
important hole in the Caspar-Klug classification~\cite{Caspar:1962a},
as it captures all known icosahedral viruses, in particular those with
an all-pentamer structure which so far defied classification. There is
also evidence that viral tiling plays a role in describing the
clustering of proteins into building blocks during capsid
assembly~\cite{Keef:2005a}. The use of group theory thus provides
important insights not easily obtained otherwise (for a more extensive
review of viral tiling,~see~\cite{Twarock:2006a}).

Apart from understanding static properties of viral capsids, one would
also like to understand the influence of symmetries on the
\emph{dynamics} of these capsids. A first step in this direction was
taken in~\cite{ElSawy:2007a}, where it was shown how the presence of
approximate inversion symmetry puts a clearly visible stamp on the
spectrum of Raman and infrared vibrational modes. This analysis
suggests that there may be various properties of capsid vibrations
which have a rather simple explanation and do not require a great deal
of knowledge about the detailed structure of the capsid.

A lot of effort has been dedicated to the study of protein vibrations
in the last forty years, and more recently on viral capsid
vibrations. The main goal so far has been to understand whether
conformational changes, which are of the utmost importance for the
proper function of these systems, occur in directions which overlap
with some low-frequency normal modes of vibration. It is however also
interesting to identify as precisely as possible the frequencies of
viral vibrations which are susceptible to be targeted by near-infrared
femtosecond laser pulses, in an attempt to mechanically destroy viral
particles without damaging neighbouring
tissues~\cite{Tsen:2007}. Despite decades of encouraging results,
almost all relying on variations of the spring-mass model and high
performance computing, the overall picture remains patchy and deserves to be
revisited.

In order to determine the spectrum of vibrations of a viral capsid,
one must, in principle, analyse a complicated interaction potential
between a large number of atoms (of the order of~$10^5$ or
$10^6$). This potential depends on bond lengths and angles, the
minimum of which describes the static configuration of the capsid.
Small vibrations are described by a harmonic potential around the
equilibrium situation. The normal modes of the associated force matrix
yield the small fluctuation spectrum of the capsid. Computer
simulations of this type have led to useful insight, about the
relation between normal and swollen forms of virus
capsids~\cite{Tama:2002a,Tama:2005a} for instance, albeit at
considerable computational cost (for recent reviews
see~\cite{Tama:2006a,Gibbons:2007a}).

It is clear that for the lowest-frequency modes, which are
characterised by slow motions with large amplitudes, many degrees of
freedom are not relevant. Various coarse-graining procedures have
therefore been proposed, among which the popular rotation-translation
block method~\cite{dura2,Tama:2005a} and the cluster normal mode
analysis~\cite{Schuyler:2005a} (see~\cite{bahar:2005a} for a review).
These have shown that, indeed, some aspects of capsid dynamics can be
understood from far fewer degrees of freedom than those present in
all-atom simulations.  Many of these developments were driven by the
requirement to bring down the substantial computational cost of
all-atom simulations.  However, even if computational cost is not an
issue, it is important to reduce the number of degrees of freedom in
order to gain insight into the \emph{systematics} of capsid dynamics.
\medskip

Instead of using the ``bottom-up'' approach described above, in which
one starts from the all-atom structural information and attempts to
coarse-grain from there, we suggest here that useful complementary
information can be obtained by using a ``top-down'' approach. The
starting point is a very minimalistic description of the capsid, in
which entire protein chains are modelled by point masses. These masses
interact with each other through springs, whose strength is determined
by the association energies of the protein chains. The relative values
for the association energies are taken from
VIPERdb~\cite{Shepherd:2006a}. This approach is similar in spirit to
the one underlying viral tiling, in the sense that we attempt to see
how much we can learn from such an extreme caricature before
introducing more degrees of freedom or more complicated interactions.

Although this drastic simplification cannot possibly be expected to
describe any but the lowest frequency modes, we will show here that it
\emph{does} uncover the systematics behind several vibration patterns
which were observed in all-atom simulations before, but for which a
simple explanation remained lacking. Further insight will be obtained
by decomposing the spectrum into irreducible representations of the
icosahedral group (a method previously used for all-atom
simulations~\cite{Vlijmen:2001,Vlijmen:2005a}).  The power of the
``top-down'' approach is thus to make it much easier to see the common
aspects of vibrational patterns among a large class of
capsids. Moreover, it allows us to systematically think about the
qualitative aspects of \emph{modifying} vibrational modes (by changing
bonds for example) without having to rely on the output of large-scale
computer simulations.

\medskip

We begin in section~\ref{s:group theory} with an overview of the role
played by group theory in the analysis of normal modes of vibration of
systems exhibiting some symmetry, and we state clearly the dynamical
hypotheses on which our modelling relies.  Section~\ref{s:alltogether}
presents the low-frequency spectra for eight viral capsids at various
$T$-numbers, namely Satellite Tobacco Mosaic, Rice Yellow Mottle,
Tomato Bushy Stunt, Cowpea Chlorotic Mottle, Polio, MS2, Hong-Kong 97
and Simian 40.  It emerges from the results obtained that the
low-frequency spectrum of all these stable\footnote{Some capsids are
  not stable when one considers the association energies given in
  VIPERdb alone. In these cases, extra theoretical bonds have been
  added to stabilise the capsid, and their occurrence is clearly
  stated in the text.} capsids~--~bar the last one, which is in a
different class, having an all-pentamer structure~--~possess 24
near-zero normal modes of vibration which always fall in the same set
of non-singlet irreducible representations of the icosahedral
group. The first singlet representation, which is associated with a
fully symmetric mode, always appears higher up in the spectrum, in
accordance with the expectation that such a motion requires more
energy to develop. We argue in section~\ref{sec-generic patterns} that
the presence of 24 near-zero-modes in the spectrum of viral capsids is
deeply rooted in the fact that the latter exhibit icosahedral
symmetry. This is done by examining the clusters of protein chains
which make up a face of the icosahedral structure, and then linking
the clusters from different faces in the minimal possible way to
obtain a stable structure. The mathematical argument is illustrated by
considering a hypothetical and very simple viral capsid, where a
single protein occupies the centre of each icosahedral
face\footnote{One may think of this approximation as being a limiting
  case for a $T=1$ capsid whose three protein chains per face are
  averaged by their centre of mass.}, and all proteins are linked to
form a dodecahedron, as was first presented
in~\cite{Englert:2008a}. More realistic capsids are then considered in
the light of the group theoretical properties discovered in the simple
case. We also offer some remarks on the role of tiling theory in our
approach to virus dynamics. Finally, we conclude with some thoughts
for future work in this direction.

\section{ Viral capsid vibrations: a group theory perspective}
\label{s:group theory}
\subsection{Symmetry-based organisation of normal modes of vibration}
\label{s:symmetry-based}

The most direct evidence of icosahedral symmetry in a large class of
viruses is the experimental observation that the proteins, which form
an almost spherical protective shell or capsid for the DNA or RNA
material, do cluster in groups of five (pentamers) around twelve
equidistant disclinations corresponding to the vertices of an
icosahedron, and in groups of six (hexamers) around points at the
intersection of the capsid with global 3-fold and/or local 6-fold
symmetry axes of the same icosahedron. This symmetry is at the core of
the Caspar-Klug classification of icosahedral viruses, which exploits
well-chosen coordinate axes on a 2-dimensional planar hexagonal
lattice to label viruses according to the number of capsid proteins
they exhibit \cite{Caspar:1962a}.

Recent theoretical work by Twarock \cite{Twarock:2004a} corroborates
the existence of an intrinsic icosahedral symmetry, even in the case
of all-pentamer capsids which appear in the polyoma- and
papilloma-viridae families. Apart from twelve pentamers located around
the global 5-fold symmetry axes of the icosahedron, more pentamers are
organised around local 5-fold symmetry axes, whose existence is deeply
rooted in the mathematics of the icosahedral group $H_3$ \footnote{The
  notation ${\cal I}_h$ is used in the chemistry literature.}. A
combination of ideas, inspired by the affinisation of this
non-crystallographic Coxeter group \cite{Patera:2003a} as well as the
theory of quasi-crystals and Penrose tilings, has led to interesting
new insights on viral capsid assembly and genome organisation
\cite{Keef:2005a,Keef:2005b,Micheletti:2006a,Keef:2008a}.

Symmetry arguments have proven very powerful in a variety of
scientific contexts, and it is natural to exploit the icosahedral
symmetry to the full in attempts to understand dynamical properties of
viruses, such as their vibration patterns. The techniques, applicable
to any complex system whose $N$ building blocks (considered as $N$
point masses) are invariant under the action of a symmetry group
$G$ \footnote{We will call such a system `a $G$-invariant, $N$-atom
  molecule' although this is obviously an abuse of language.}, date
back to Wigner \cite{Wigner:1930} and have been applied in chemistry
as early as 1934 \cite{Wilson:1934}. Since then, the frequencies of
vibration of the most standard chemical molecules, and of some
fullerene structures, have been analysed with group theory methods in
the approximation where the potential is quadratic.

Remarkably, many qualitative features of the normal modes of vibration
depend mainly on the symmetry of the system, and not on the potential
chosen. For instance, it is quite straightforward to decompose the
motions of a `$G$-invariant $N$-atom molecule' into irreducible
representations of the symmetry group $G$ without reference to the
harmonic potential chosen to model interactions between the
atoms. Such decomposition provides further insight into the
universality of certain motions, and we emphasise that it does
\emph{not} imply that we only keep the modes which respect the
symmetry $G$. The potential is required to be G-invariant, but the
solutions to the equations of motion, i.e.~the vibration modes, can
break this symmetry.

The `molecule' we are interested in is a viral capsid, whose `atoms'
are its $N$ capsid proteins, each of them approximated by a point mass
located at its centre of mass. The symmetry group $G$ is the
icosahedral group $H_3$ if the capsid exhibits a centre of inversion,
or its 60-dimensional proper rotation subgroup ${\cal I}$ otherwise.
We remark here that Nature does not seem to favour viruses whose
capsid proteins have a centre of inversion at the atomic
level. Nevertheless, in coarse-grained approximations, including the
one used here, the configuration of the $N$ point masses chosen to
model the capsid may be very close to having a centre of
inversion. Consequences of this remark were explored in
\cite{ElSawy:2007a} in the context of viral tiling theory.

The decomposition of the vibrational modes of such a capsid into
irreducible representations of the group $H_3$ (resp.~${\cal I}$) is
standard and is sketched in appendix~\ref{app-patterns}. For the
proper rotation subgroup ${\cal I}$ of the icosahedral group, the
generic decomposition of the displacement representation (which
encodes the action of ${\cal I}$ on the displacements of the $N$
protein-point masses from their equilibrium position due to
vibrations) reads,
\begin{equation}
\label{e:3Ndecomp}
\Gamma^{3N}_{\text{displ}} = \frac{3\,N}{60}\Big[ \Gamma^{1} + 3
  \Gamma^{3} + 3 \Gamma^{3'} + 4\Gamma^{4} + 5\Gamma^{5}\Big]\,,
\end{equation}
where all representations $\Gamma$ on the r.h.s. are irreducible, and
the numerical superscripts refer to the dimension of the
representations. The decomposition~\eqref{e:3Ndecomp} does not contain
any information about the force matrix, introduced in
appendix~\ref{app-force}, and hence does not provide any insight into
the geometry of the vibration modes. In order to obtain that
information, while preserving the group theoretical structure, one
must turn to a block diagonalisation of the force matrix, as
illustrated in appendix~\ref{app-patterns} in the case of the ammonia
molecule. Such a detailed analysis reveals that, in accordance with
the decomposition \eqref{e:3Ndecomp}, the capsid undergoes vibrational
motions which can be expressed in terms of a linear combination of
independent normal modes, some of them degenerate, in the sense that
they all have the same frequency and are transformed into each other
under the action of ${\cal I}$. The expression \eqref{e:3Ndecomp}
encodes how the $3N$ normal modes of vibration are organised in
subsets (irreducible representations) such that all normal modes
within a given subset have the same frequency and transform into each
other under the group action. The number of elements in a subset is
given by the dimension on the corresponding irreducible
representation.

Not all $3N$ normal modes of vibration of our viral capsids are
`genuine'. Indeed, in this 3-dimensional problem, three degrees of
freedom correspond to the translations, and three to the rotations of
the capsid as a whole. These zero-modes belong to two $\Gamma^{3}$
irreducible representations and are usually discarded from dynamical
considerations as they are trivial.

Further standard group theory considerations provide an easy method to
pin down which are, among the normal modes described
in~\eqref{e:3Ndecomp}, those one could in principle detect using Raman
and infrared spectroscopy. The former relies on a physical phenomenon
induced by sending a beam of frequency $\nu$ on a molecule or capsid:
some radiation is scattered, whose frequency is the incident frequency
$\nu$ shifted by $\pm \nu_s$. The electric field $\vec{E}$ carried by
the incident beam induces a dipole moment $\vec{\mu}$ in the molecule
or capsid, such that
\begin{equation}
\mu_i=\alpha_{ij}E_j,\qquad i, j =1,2,3\,,
\end{equation}
and the polarisability tensor $\alpha_{ij}$ transforms under the
symmetry group of the molecule as the six quadratic expressions
$x_ix_j$, where $x_i, i=1,2,3$ are the coordinates of a point in
space. For the group ${\cal I}$, the quadratic expression
$x_1^2+x_2^2+x_3^2$ transforms as a singlet (and thus belongs to
$\Gamma^{1}$) while the other five independent quadratic expressions
belong to $\Gamma^{5}$. From~\eqref{e:3Ndecomp}, one thus concludes
that our $N$ protein-point mass capsid has $3N/60$
non-degenerate\footnote{The number $N$ of capsid proteins of
  Caspar-Klug viruses is a multiple of 60.}  and $N/4$ five-fold degenerate
Raman active modes of vibration.  On the other hand, infrared active
modes belong to $\Gamma^{3}$ irreducible representations (see
\cite{Bishop:2003} for instance), and therefore the capsid possesses
$(9N/60 -2)$ three-fold degenerate infrared active modes. Such information
might become useful when experiments will be sensitive enough to
measure very low-frequency modes in macro-biomolecular assemblies.

\subsection{Overview of existing models}
\label{s:overview}

Normal mode analysis (NMA) has been quite successful in its attempts
to describe the conformational changes in a variety of
proteins~\cite{Karplus:1976, Noguti:1982a, Brooks:1983a, Go:1983,
  Levitt:1983, Harrison:1984}, and it proves a useful tool in the
study of the dynamics of large macro-biomolecular assemblies, in
particular viruses. In this context, the idea is to verify whether
various experimentally observed conformations of a given viral capsid
could be inferred from each other by arguing that conformational
changes occur in directions which maximally overlap with those of a
few low-frequency normal modes of the capsid~\cite{
  Simonson:1992,Tama:2002a,Vlijmen:2005a, Rader:2005a}.

The method, however, has limitations. One should keep in perspective
that strictly speaking, biologically significant low-frequency motions
are typically not vibrational, due to the damping influence of the
environment. Furthermore, NMA assumes the existence of a single well
potential whose minimum is a given stable configuration of the viral
particle studied, overlooking the possibility of neighbouring
multi-minima of energy. Also, the harmonic approximation to the
single-well potential most analyses consider is only valid if the
particle undergoes small motions, and this does not lend itself to an
accurate description of the observed conformational
changes. Nevertheless, NMA provides dynamical data which are
consistent with experimental results, especially on proteins
\cite{Brooks:1985, Gibrat:1990, Marques:1995, Mouawad:1996}, and is
supported by a recent statistical study \cite{Alexandrov:2005}. It
thus seems reasonable to continue to use the method, provided the
results are interpreted in the light of the caveats above.

Viruses are much more complex structures than proteins, and the
biggest challenge remains the choice, within the NMA framework, of a
potential which captures the physics of capsid vibrations whilst
taking into account a reduced number of degrees of freedom to enable
practical calculations.  Many NMA applied to viruses implement
variations of the simple Elastic Network Model proposed a decade
ago~\cite{Tirion:1996a}, in which the atoms are taken as point masses
connected by springs modelling interatomic forces, provided the
distance between them is smaller than a given cutoff
parameter. Simplified versions include the restriction to
$C^{\alpha}$-atoms only, the approximation in which each residue is
considered as a point mass, or where even larger domains within the
constituent coat proteins are treated as rigid blocks
\cite{Tama:2002a}.

The elastic potential in all analyses above has two major drawbacks:
it does not discriminate between strong and weak bonds since it
depends on a single spring constant, and it uses the rather crude
technique of increasing the distance cutoff to resolve capsid
instabilities. Consequently, the frequency spectra have much less
structure than one would expect in reality, and in particular fail to
reproduce areas of rigidity and flexibility of the capsid
satisfactorily. This phenomenon is illustrated in
appendix~\ref{app-Tirion}, where a system of 8 atoms is studied from
two different perspectives. On the one hand, the system is viewed as
two proteins consisting of four atoms each, with interactions
characterised by two different bond strengths. On the other hand, the
system of 8 atoms is subjected to a Tirion potential with varying
cutoff distance but using only one bond strength. The former is
analysed within the rotations-translations of blocks method
(RTB)~\cite{dura2} and provides a spectrum with enhanced structure,
when set against the Tirion-based analysis.

In order to obtain more accurate spectra, the authors
in~\cite{Kim:2006a,Kim:2006b,Kim:2006c} implement a bond-cutoff
method. An elastic network whose representatives are $ N\,
C^{\alpha}$- atoms is set up such that four consecutive
$C^{\alpha}$-atoms are connected via springs, introducing $3N-6$
constraints in the system.  This backbone modelling provides stability
of proteins with a less intricate elastic network than the one
obtained via the distance-cutoff method. Further springs with
different spring constants are added to model the various types of
chemical interactions (disulfide bonds, hydrogen bonds, salt-bridges
and van der Waals forces) within each protein. The proposed model
reproduces conformational changes better than the conventional
distance-cutoff simulations. Adapting this model to viruses would
certainly be enlightening, but remains a computational challenge at
present. Our model is close in spirit to the above, but takes fewer
degrees of freedom into consideration. It should be seen as a first step
towards an implementation of the programme
in~\cite{Kim:2006a,Kim:2006b,Kim:2006c} for viral capsids.

\subsection{Dynamical hypotheses in our model}

We model capsids ``top-down'', that is, by starting from as few
degrees of freedom as possible. A minimal set consists of the
centre-of-mass positions of the protein chains, as well as sufficient
bonds to make the capsid stable. For the equilibrium positions of the
proteins we make use of the data in VIPERdb~\cite{Shepherd:2006a}. The
equilibrium positions of the protein chains are assumed to respect
icosahedral symmetry.

The inter-protein forces away from equilibrium are approximated by
a harmonic potential, but we allow for the spring constants to be
different for every bond. Denoting the spring constant between protein~$n$
and $m$ by $\kappa_{mn}$, we thus have
\begin{equation}
\label{e:springmassV}
V = \sum_{\stackrel{m<n}{ m,n=1}}^N \frac{1}{2}\kappa_{mn} \Big( 
 | \vec{x}_{m}-\vec{x}_{n} | - |\vec{x}^{\,0}_{m}-\vec{x}_{n}^{\,0}|\Big)^2\,.
\end{equation}
Here~$\vec{x}_m$ denotes the actual position of the protein and
$\vec{x}_{m}^{\,0}$ its equilibrium position. For small deviations from
equilibrium, we can expand the potential as
\begin{equation}
V = \sum_{m=1}^N \left.\frac{\partial V}{\partial x^i_m}\right|_{x=x^0} (x_m^i-x^{0i}_m)
+ \sum_{m,n=1}^N\frac{1}{2}\left.\frac{\partial^2 V}{\partial x^i_m x_n^j}\right|_{x=x_0}
   (x_m^i-x^{0i}_m)(x_n^j-x^{0j}_n) + \ldots\,.
\end{equation}
As there is only very little spread in the values of the protein
masses in virus capsids, we will normalise them all to one, thereby
absorbing the overall mass into the spring constants~$\kappa_{mn}$.
The equations of motion for the deviations from equilibrium then
become
\begin{equation}
\frac{{\rm d}^2}{{\rm d}t^2} \big( x_m^i - x_m^{0i}\big)
 + F_{mn}^{ij} \big(x^j_n - x_n^{0j}\big) = 0\,,\qquad i,j =1,2,3\,,
\end{equation}
where the force matrix is obtained as the second derivative of the
potential with respect to the positions, evaluated at the equilibrium
positions. Explicitly, using the potential~\eqref{e:springmassV}, one finds
\begin{equation}
\label{e:force_matrix}
F^{ij}_{mn} = \left.\frac{\partial^2 V}{\partial x_m^i \partial x_n^j}\right|_{x=x^0}
= \begin{cases}\displaystyle
\left.\sum_{p\not=m} \kappa_{mp} \frac{(x_m-x_p)^i (x_m-x_p)^j}{(x_m -
  x_p)^2}\right|_{x=x^0} & \text{if~$m=n$}\,,\\[2ex]
\displaystyle -\kappa_{mn} \left.\frac{(x_m-x_n)^i (x_m-x_n)^j}{(x_m -
  x_n)^2}\right|_{x=x^0}& \text{otherwise}\,.
\end{cases}
\end{equation}
It is convenient to store all degrees of freedom in one vector,
i.e.~to introduce a vector~$\vec{q}$ with components given by $q^{3m -
  3 + i} = r_{m}^{i}\equiv x_m^i-x_m^{0i}$. In terms of this vector
the equation of motion reads
\begin{equation}
\label{e:EOM}
\ddot{\vec{q}} + F \vec{q} = 0\,,
\end{equation}
and the frequencies of the normal modes are given by the eigenvalues
of the matrix~$F$, as reviewed in~\ref{app-force}.  Note that this
matrix depends explicitly on the equilibrium positions of the masses.

The determination of the frequencies is therefore straightforward in
principle. However, given the fact that one is faced with highly complex
systems, the rapidly increasing number of degrees of freedom and the
shape of the potential put severe constraints on actual
calculations. Although results based on a truly phenomenological
potential are still out of reach, our model nevertheless paves the way
to more realistic situations. It introduces a drastic coarse-graining
but on the other hand, provides a hierarchy between bonds, dictated by
the relative values of association energies listed in VIPERdb. It also
explores the conditions for iso-staticity of the viral capsids, in the
light of the icosahedral symmetry inherent to the system. In other
words, the model is simple enough to enable the identification of
which type of bonds are crucial or not for the stability of the
capsids, as will be developed in section~\ref{sec-generic patterns}.

Beforehand, we calculate in section~\ref{s:alltogether} the
low-frequency spectra of a variety of Caspar-Klug capsids within the
technical framework summarised in the present section, and highlight
the remarkable feature announced in the Introduction, namely the
presence of a very low-frequency plateau of twenty-four normal modes
in all Caspar-Klug viruses studied. We also analyse SV40, an
all-pentamer capsid which falls out of the Caspar-Klug classification,
and note that the nature of the plateau is slightly different. A
proper understanding of this phenomenon requires further investigation
and is beyond the scope of this paper.

\section{Results for specific virus capsids}
\label{s:alltogether}

In order to explore the consequences of our dynamical hypotheses we
have analysed a variety of virus capsids. Their Caspar-Klug
$T$-numbers range from 1 to 7, and we have paid particular attention
to $T=3$ capsids in order to assess the role of viral tiling theory in
shaping vibrational spectra. We have indeed chosen our~$T=3$ capsids
so that the 3~tiling types (triangle, rhomb and kite) compatible with
the structure of 12 pentamers and 20 hexamers are
represented. Table~\ref{t:summarytable} summarises our choice of
capsids. We now discuss each capsid separately and highlight various
interesting features as we go along.

%

\begin{table}[ht]
\vspace{2ex}
\begin{center}
\begin{tabular}{llllllr}
name               & abbr. & pdb      & $T$   & approx.~centre   & tiling
& zero \\
                   &       & code     &       & of inversion &
& modes
 \\[1ex]
\hline\\[-1ex]
Satellite Tobacco Mosaic  & STMV  & 1A34     & 1   & no          & triangle & 6 \\[2ex]
Rice Yellow Mottle        & RYMV  & 1F2N     & 3   & no          & triangle & 6 \\
Tomato Bushy Stunt        & TBSV  & 2TBV     & 3   & no          & rhomb    & 30 \\
Cowpea Chlorotic Mottle   & CCMV  & 1CWP     & 3   & no          & rhomb    & 90 \\
Polio                     & Polio & 2PLV     & 3   & no          & kite     & 6 \\
MS2                       & MS2   & 2MS2     & 3   & yes         & rhomb    & 90\\[2ex]
Hong Kong '97             & HK97  & 2FTE     & 7$l$  & no        & rhomb    & 6 \\
Simian 40                & SV40  & 1SVA     & 7$d$  & yes       & rhomb/kite & 6
\end{tabular}
\end{center}
\caption{Summary of the viruses analysed in this section, together
  with some of their fundamental properties. If the number of zero
  modes is larger than~6 additional bonds, over and above those given
  in VIPERdb, have to be added to stabilise the capsid.\label{t:summarytable}}
\end{table}

\subsection{Satellite tobacco mosaic virus}

The STMV virus capsid is one of the simplest capsids to analyse since
it has $T=1$ and only 60 capsid proteins. An all-atom molecular
dynamics simulation for STMV was reported on
in~\cite{Freddolino:2006a}. One of the main conclusions of their
analysis is that the empty capsid (i.e.~without RNA content) is
unstable. The instability manifests itself in the behaviour of two
non-adjacent faces around a 5-fold symmetry axes, which sink into the
interior~\cite{Freddolino:2006a}. Although a molecular dynamics
simulation of this type cannot reach the very lowest frequency modes
because the maximal evolution time is limited, we can still use this
analysis to draw some general conclusions about the low-frequency
spectrum. An important result is that the proteins remain essentially
undeformed. This yields support to one of our dynamical hypotheses,
namely that the individual atoms from which the proteins are built can
be grouped together for an adequate analysis of the low-frequency
modes. The collapse should thus be part of the low-frequency spectrum,
as captured by a coarse-grained model.

\begin{figure}[t]
\begin{center}
\includegraphics[height=.3\textwidth]{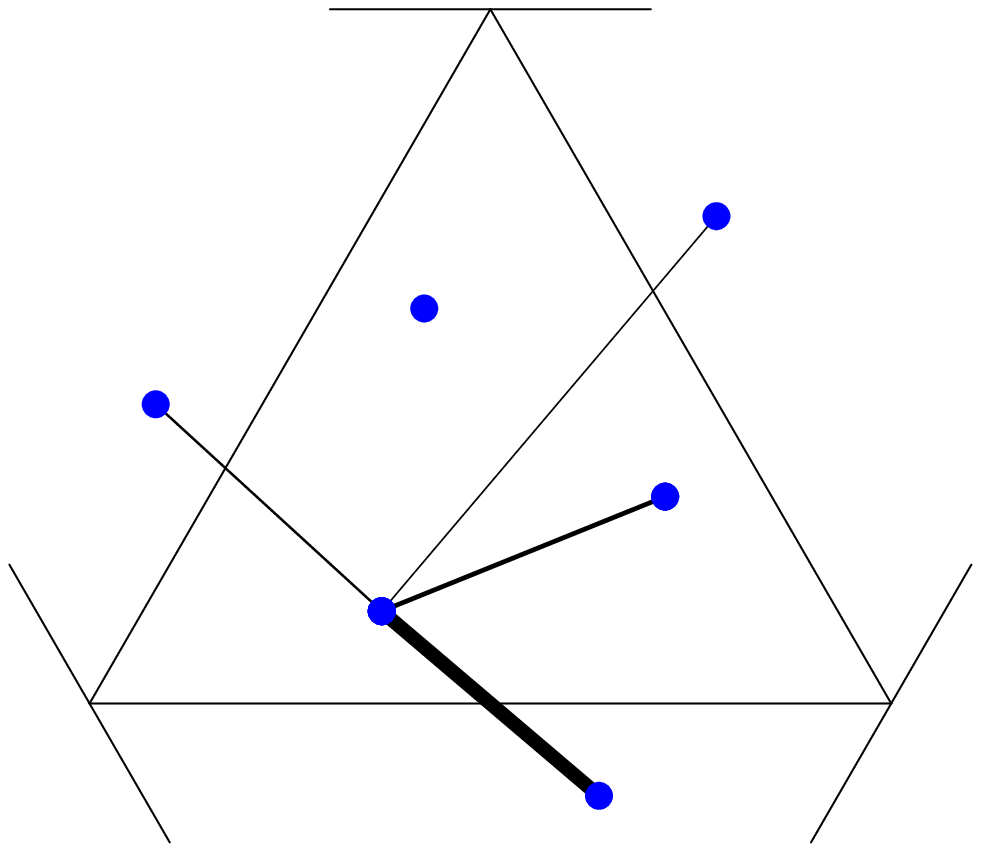}
\qquad
\includegraphics[height=.3\textwidth]{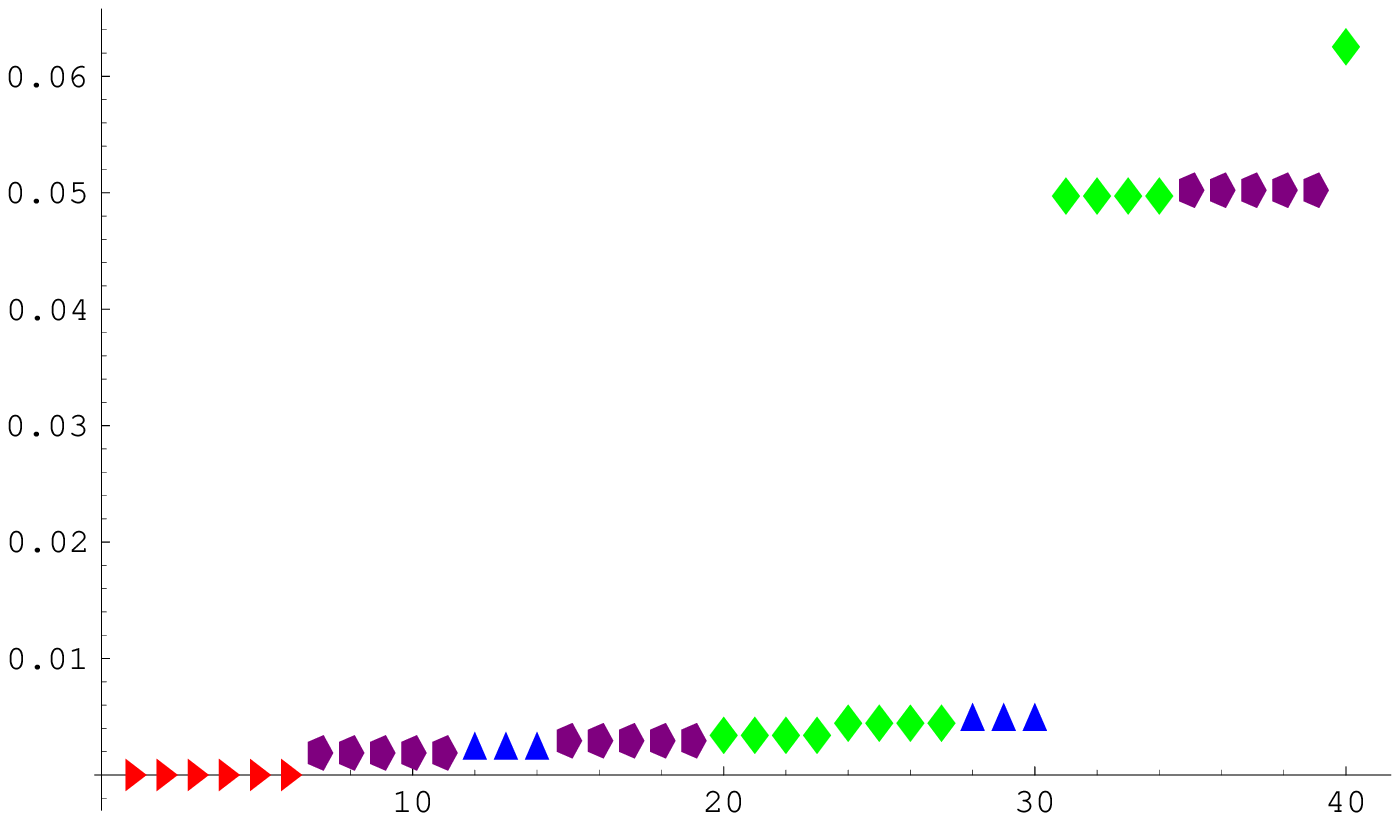}
\end{center}
\vspace{-1ex}
\caption{The STMV protein positions and the inter-protein bonds (left).
  The longest bond in the figure is the weak A1-A8 bond.  Displayed on
  the right is the spectrum of the 40 lowest-frequency normal modes
  of STMV (up to an overall normalisation). Clearly visible are the 6
  trivial zero-frequency modes as well as the 24 low-frequency
  modes. The modes $\rhd$ (resp.~$\triangle$) belong to 3-dimensional
  irreducible representations $\Gamma^3$ (resp.~$\Gamma^{3'}$)
  of the icosahedral group. The diamond (resp.~pentagon) modes belong
  to 4 (resp.~5)-dimensional irreducible representations. The $x$-axis
  labels the normal modes while the $y$-axis gives the wave numbers up
  to an overall normalisation.\label{f:STMV_fund_freq}}
\end{figure}

In our simplified model the capsid is actually stable, but it has 24
normal modes with relatively low frequency. The bond structure,
obtained from the association energies as described earlier, is
displayed in \figref{f:STMV_fund_freq}. The resulting low-frequency
spectrum is displayed there as well, with modes marked according to
their representation content. The most significant feature of this
spectrum is the appearance of a low-frequency plateau of 24~modes,
separated by a large gap from the remainder of the spectrum. The
representation content of the plateau is
\begin{equation}
\label{e:STMV_plateau}
2\Gamma^{3'} + 2\Gamma^{4} + 2 \Gamma^{5}\,.
\end{equation}
Within our model, it is easy to verify that the height of this plateau
is related to the strength of the long A1-A8 bond. In the data
provided by VIPERdb, this bond has a strength of only about 10\% of the
strongest bond present. By removing the A1-A8 bond, the
24~low-frequency plateau modes come down to zero frequency.

If one focuses only on the structure on each individual icosahedral
face, one observes that the three protein chains form a relatively
rigid triangle, which has 6~zero-modes (three rotations and three
translations). Without the A1-A8 bond, there are 3 edge-crossing bonds
per edge, as can be reconstructed from \figref{f:STMV_fund_freq} by
using the 2, 3 and 5-fold symmetries of the icosahedron.  Since each
bond induces one constraint in the system, naive counting suggest that
there should be $6\cdot 20 - 3\cdot 30 = 30$ zero-modes in this case,
as the icosahedron has 20 faces and 30 edges. We will see how this
type of reasoning provides insight in many other virus capsids in
section~\ref{s:faces}.

\subsection{Rice yellow, tomato bushy stunt and cowpea chlorotic
  mottle virus}
\label{s:RYM_TBSV_CCMV}

The RYMV, CCMV and TBSV capsids share many features, some of which are
obvious from their common~$T=3$ structure, while others are less
manifest. One of the most manifest differences is their viral tiling
structure. While RYMV has a triangle tiling, both TBSV and CCMV are
tiled by rhombs. The difference in tiling leads to a difference in
dominant bond structure, but as we will see, this turns out to be of
relatively little importance for the low-frequency spectrum. A
somewhat related~$T=3$ virus, Polio, has an additional protein chain
hidden slightly inside the main capsid. This kite-tiled virus will be
discussed separately.

\begin{figure}[t]
\begin{center}
\includegraphics[height=.3\textwidth]{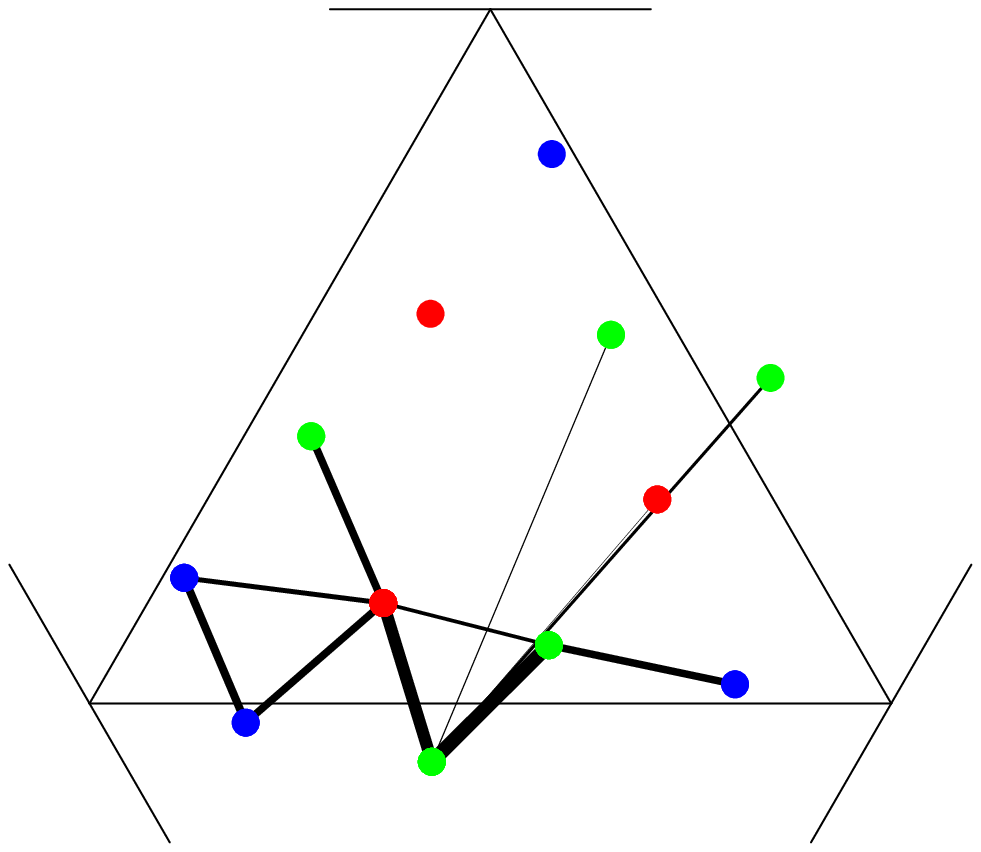}
\qquad
\includegraphics[height=.3\textwidth]{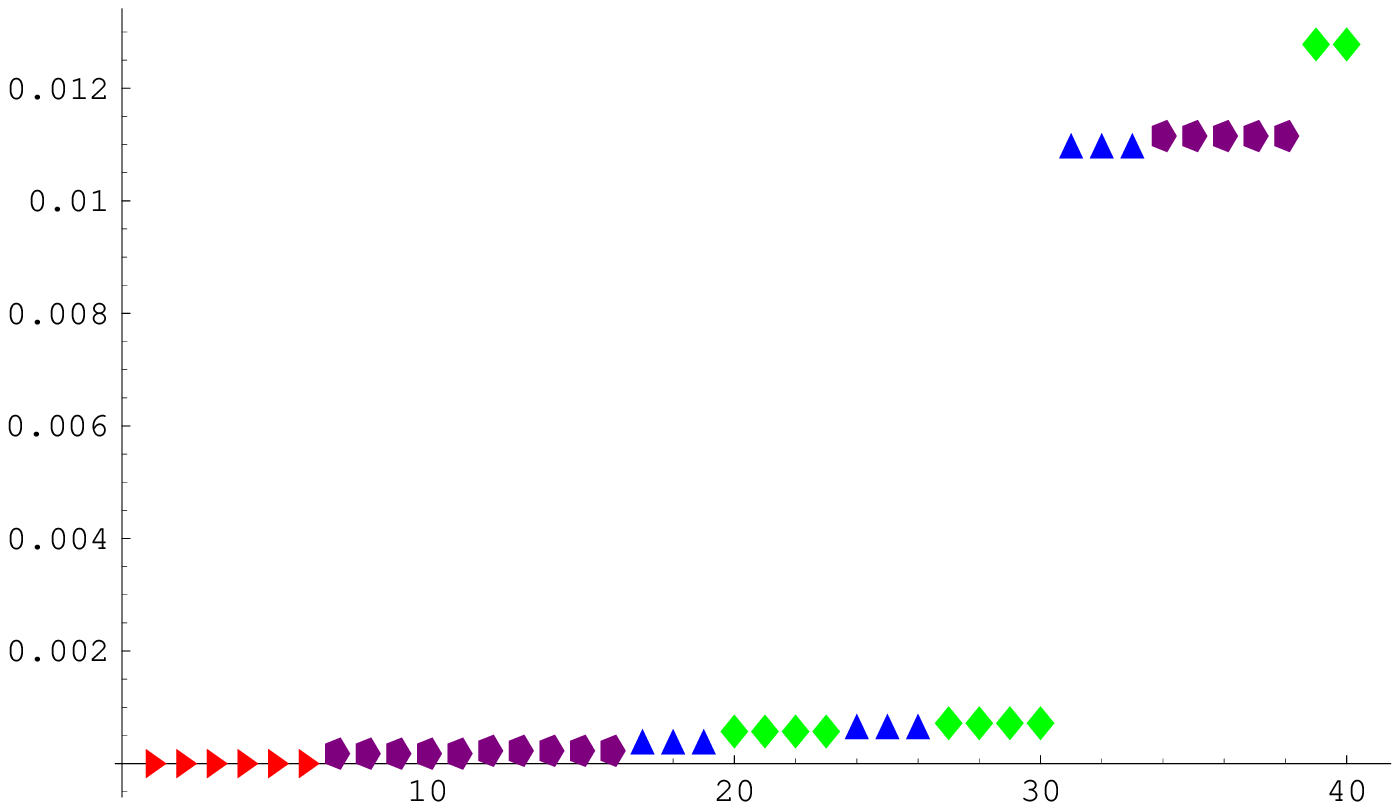}
\end{center}
\vspace{-1ex}
\caption{The protein positions and inter-protein bonds for RYMV (left), as
  well as the 40 lowest-lying normal mode frequencies (right). Again, there are 6
trivial zero-modes and 24 low-frequency modes, after which the
frequencies go up rapidly.\label{f:1F2N_fund_freq}}
\vspace{2ex}
\begin{center}
\includegraphics[height=.3\textwidth]{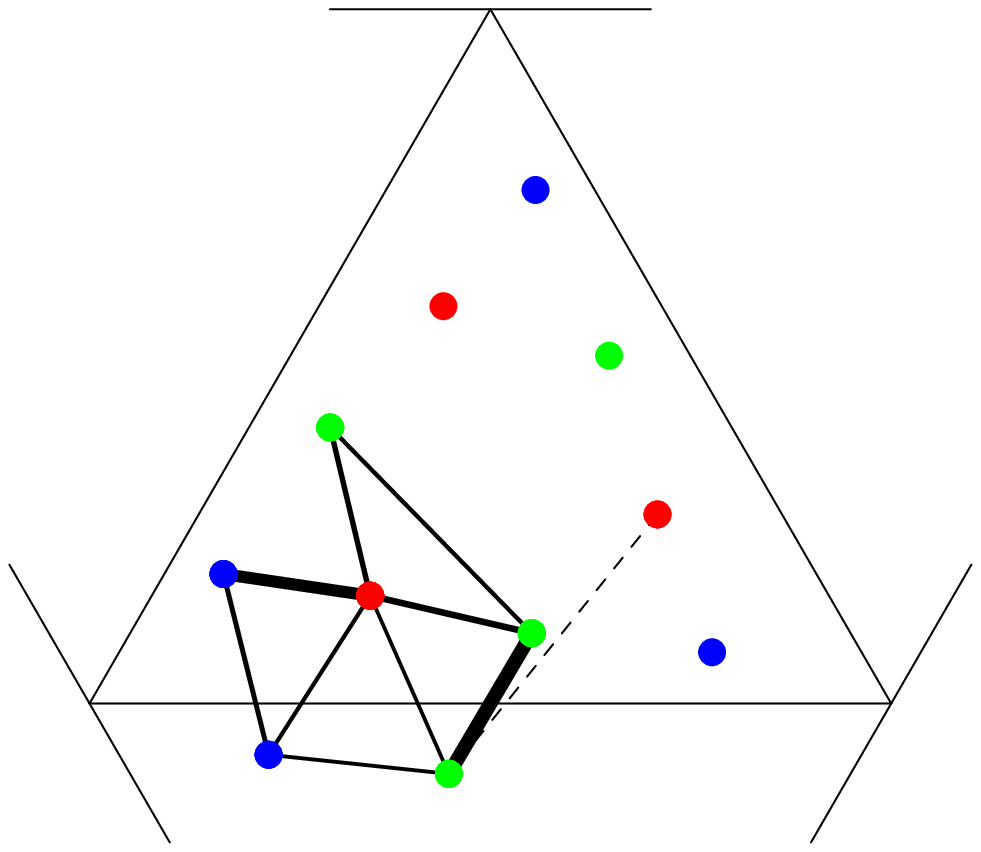}
\qquad
\includegraphics[height=.3\textwidth]{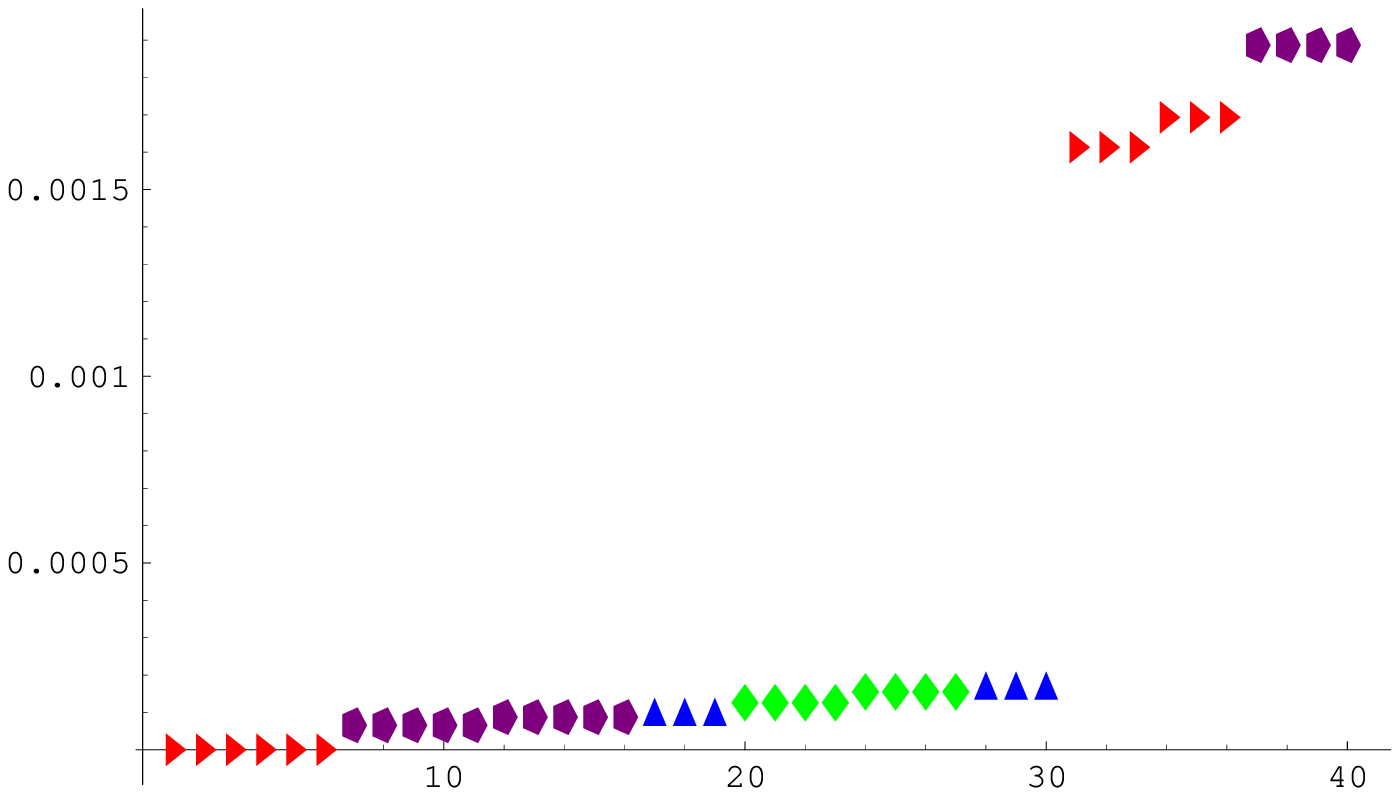}
\end{center}
\vspace{-1ex}
\caption{The protein positions and inter-protein bonds for TBSV
  (left), as well as the 40 lowest-lying normal mode frequencies
  (right). The bond indicated with a dashed line is a very weak bond
  which was added by hand to make the capsid
  stable.\label{f:2TBV_fund_freq}}
\end{figure}

Let us start with RYMV, which has a triangle tiling. Its spectrum was
computed using RTB techniques in~\cite{Tama:2005a} by making use of a
Tirion potential with $C^\alpha$ atoms as fundamental degrees of
freedom. Their analysis, however, is focused on the icosahedrically
symmetric (non-degenerate, singlet) modes, while quite a few
non-symmetric (degenerate) modes have lower frequency. In our model,
employing VIPERdb association energy bonds, the capsid is stable and
meaningful frequencies can thus be extracted. The spectrum of the
first 40 modes is displayed in \figref{f:1F2N_fund_freq}. Just as for
the simpler STMV capsid discussed above, the most manifest feature is
a plateau of 24 low-frequency modes (apart from the 6~trivial rotation
and translation modes). This plateau in fact has precisely the same
representation content as the one found for STMV,
see~\eqref{e:STMV_plateau}.

\begin{figure}[t]
\begin{center}
\includegraphics[height=.4\textwidth]{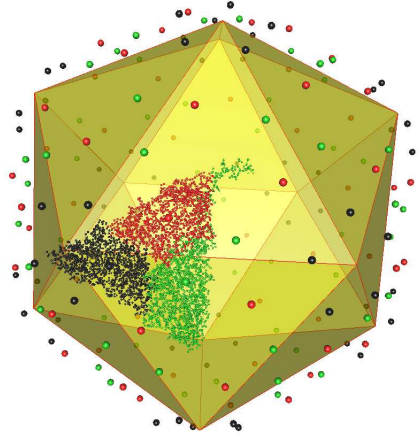}
\end{center}
\vspace{-1ex}
\caption{The capsid of RYMV. The large dots represent centre-of-mass
  positions of protein chains, while the small dots represent
  individual residues for three of the chains. Clearly visible is the
  long arm of the (green) C-chain which passes to the right of the
  (red) B-chain and stretches towards the top-right of the figure to
  establish the capsid-stabilising bonds.\label{f:RYM_C-chain}}
\end{figure}

The bond structure of RYMV exhibits, apart from ``nearest neighbour''
interactions, three long-range bonds which stretch out from the
C-chains (see \figref{f:1F2N_fund_freq}; one of these arms
stretches from a (green) C-chain to a (red) B-chain and is hidden from
view). These bonds are a consequence of the long tail of atoms which
extends from the C-chain (see \figref{f:RYM_C-chain}), and turn
out to be crucial for stability of the capsid. Artificially removing
the weakest two introduces additional zero-modes which signal an
instability against deformation. More precisely, what happens is that
removal of these long-range bonds brings down the plateau of 24
low-frequency modes, turning them into zero-modes. Again, this is a
perfect analogy with what happens for STMV.

Turning now to TBSV, we first of all note that the association
energies given in VIPERdb are not sufficient to make the capsid
stable. In fact, the TBSV spectrum has, in our model, a total of 30
zero-modes (including the trivial ones). By introducing one additional
weak bond (indicated by a dotted line in \figref{f:2TBV_fund_freq}),
it turns out that 24~of these zero-modes get lifted. The resulting
spectrum is again of the type we have seen before: a low-frequency
24-state plateau~\eqref{e:STMV_plateau}, followed by a gap.

\begin{figure}[t]
\begin{center}
\includegraphics[height=.3\textwidth]{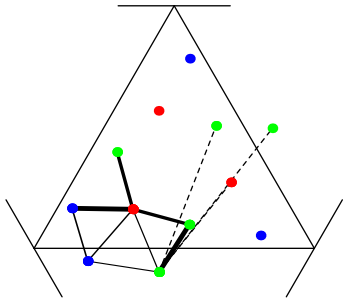} 
\end{center}
\vspace{-4ex}
\caption{The protein positions and inter-protein bonds for CCMV as
  given in VIPERdb. The dashed bonds were added by hand to stabilise the
  capsid.\label{f:CCMV_fund}}
\end{figure}

Let us finally discuss CCMV. Its protein positions are similar to
those of RYMV, but the bonds suggest a rhomb tiling instead of a
triangular one. Unfortunately the association energies listed by
VIPERdb are not sufficient to make the capsid stable, but we can add
three weak long-range bonds similar to those in RYMV for stabilisation
(see \figref{f:CCMV_fund}). Given the difference in structure of the
strong bonds (as compared to RYMV), one might now naively expect a
rather different low-frequency spectrum. However, it turns out that
CCMV again exhibits the by now familiar low-frequency 24-state plateau
followed by a rather large gap (the size of which depends on the
strength of the bonds which were added by hand). It is at present not
easy to compare these results with other studies of the CCMV
capsid. The analysis of~\cite{Tama:2005a} is based on a
$C^\alpha$-atom analysis using a Tirion potential. As we argue in
appendix~\ref{app-Tirion}, this potential has the tendency to smooth
out sharp features of the spectrum. The fact that~\cite{Tama:2005a}
find a singlet mode already after 17 non-singlet low-frequency modes
may thus be due to this difference in the potential used. Dynamical
aspects of CCMV have also been analysed by~\cite{Hespenheide:2004a},
who focused on the connectivity properties of the capsids rather than
the precise form of the force matrix. Given the restrictions of the
VIPERdb bonds, we will refrain from making a comparison with their
results at this stage.

\subsection{Polio virus}

\begin{figure}[t]
\begin{center}
\includegraphics[height=.3\textwidth]{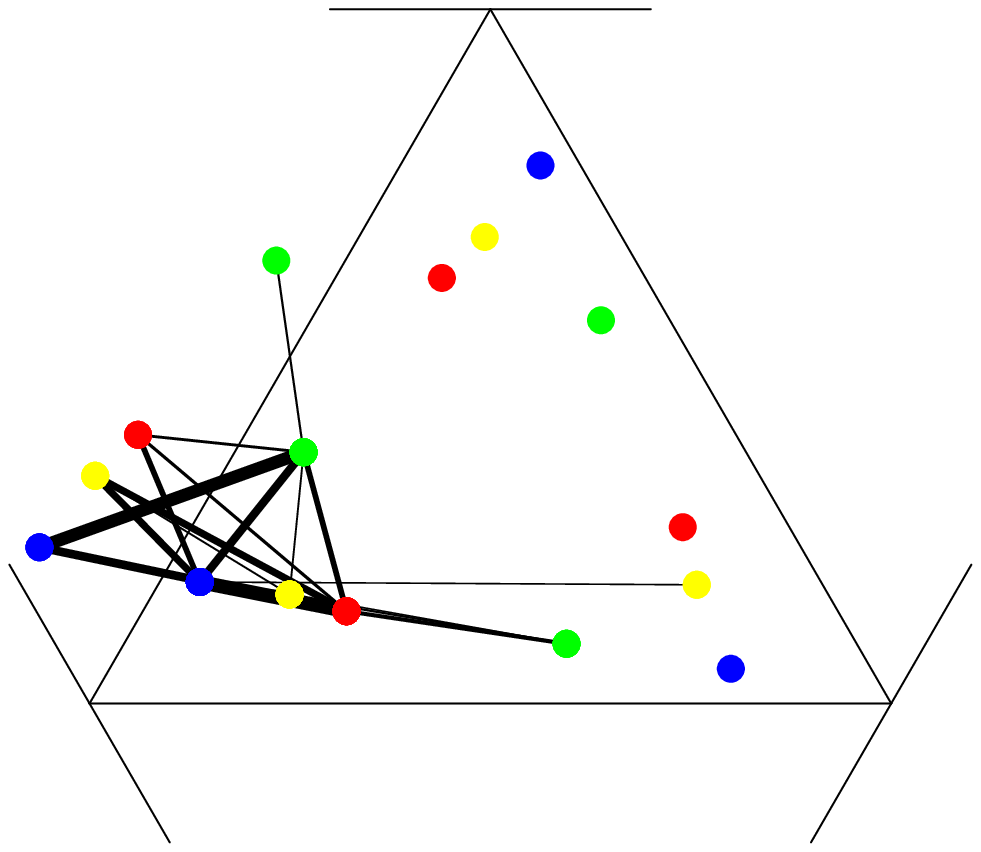} \qquad
\includegraphics[height=.3\textwidth]{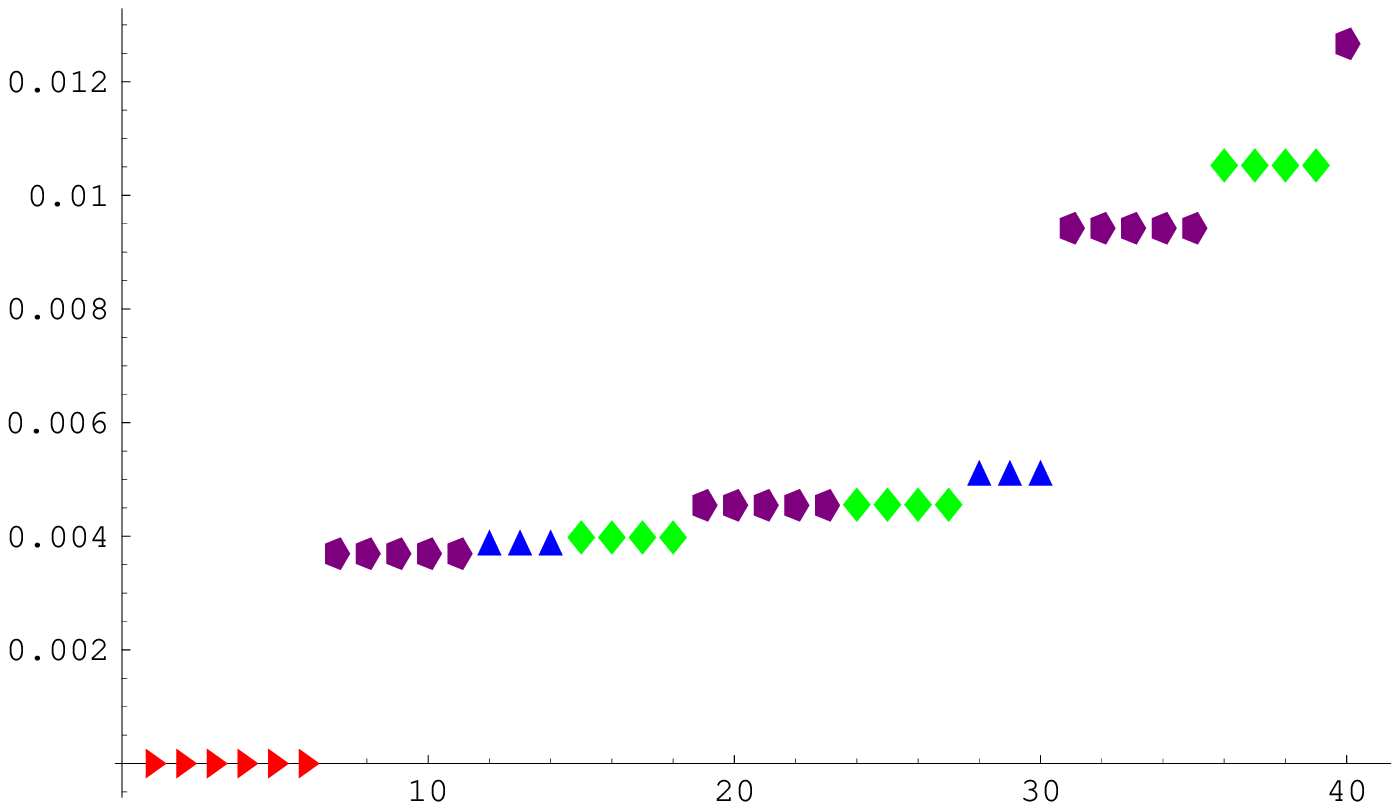}
\end{center}
\vspace{-1ex}
\caption{The protein positions and inter-protein bonds for the Polio
  virus (left), as well as the lowest-lying normal mode frequencies
  (right). Again, there are 6 trivial zero-modes and 24 low-frequency
  modes, after which the frequencies go up
  rapidly.\label{f:POL_fund_freq}}
\end{figure}

The polio virus is a pseudo~$T=3$ virus. It has four protein chains,
one of which lies inside the main capsid. Ignoring this fourth chain,
the tiling suggested by the remaining 180 protein positions is of kite
type. The spectrum obtained from our model is displayed in
\figref{f:POL_fund_freq}. Once again, it exhibits a low-frequency
plateau of 24 modes with the representation
content~\eqref{e:STMV_plateau}. The lowest frequency modes, and in
particular the characteristic motions of the proteins, have also been
analysed to some extent in~\cite{Vlijmen:2005a}. Of particular
interest to us are their results on the representation content of the
lowest frequency modes, which differs slightly from ours, and a
visualisation of the motion of the atoms. The origin of this
discrepancy remains unclear, as the latter paper is not very specific
about the potential function used.

\begin{figure}[t]
\begin{center}
\includegraphics[height=.3\textwidth]{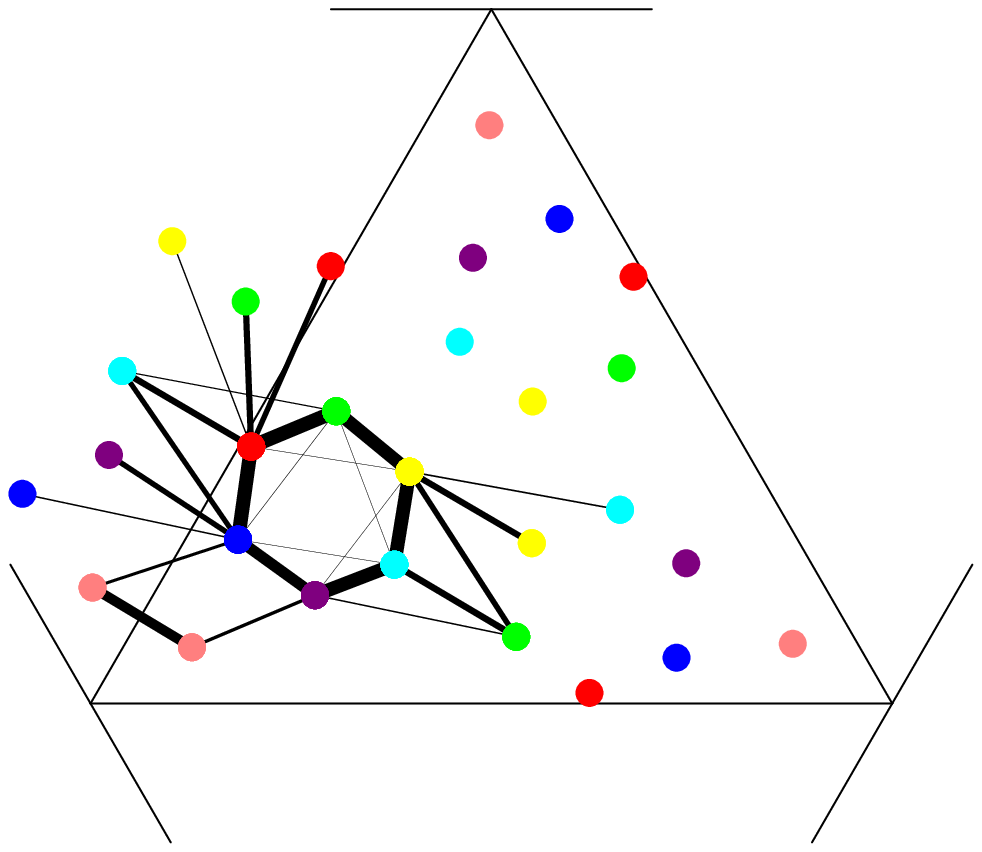}
\qquad
\includegraphics[height=.3\textwidth]{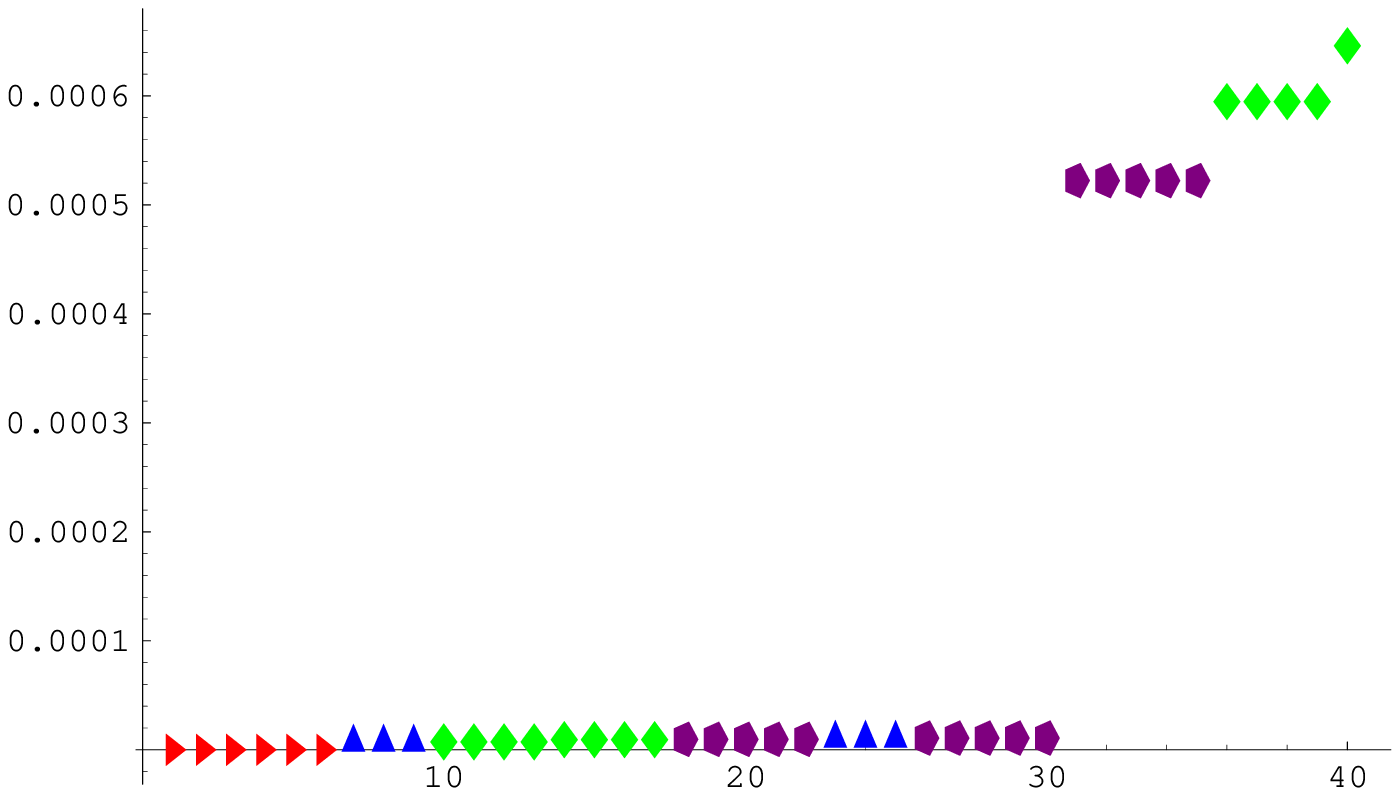}
\end{center}
\caption{The protein positions and inter-protein bonds for HK97 (left)
  as well as the low-frequency spectrum (right).\label{f:HK97_fund_freq}}
\end{figure}

\subsection{MS2}

\begin{figure}[t]
\begin{center}
\includegraphics[height=.3\textwidth]{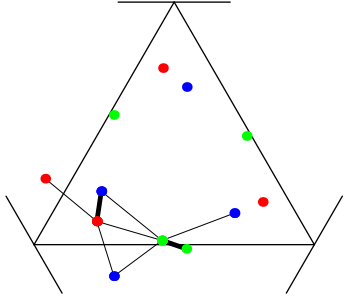}
\end{center}
\caption{The protein positions and inter-protein bonds for MS2.\label{f:MS2_fund}}
\end{figure}

The inter-protein bonds for MS2 as given by VIPERdb are not sufficient
to ensure stability of its capsid. For completeness we display the
known bonds in \figref{f:MS2_fund}. Although the number of zero-modes
(90) is the same as for CCMV, we now no longer have access to a
similar but stable capsid which we can use as a guideline to add
stabilising bonds by hand (as we had for CCMV, where we used the
similarity with RYMV). We therefore refrain from discussing the
low-energy spectrum of MS2 here. It would be interesting to revisit
the computation of the association energies for MS2, or alternatively
get a better handle on the inter-protein bonds of this virus directly from
experiment.

\subsection{Hong Kong '97}

The Hong Kong '97 virus has a~$T=7l$ capsid. Its hexamer rings are
bound rather rigidly, as shown in \figref{f:HK97_fund_freq}. Once
more we find that the low-frequency spectrum is dominated by a
24-state plateau, clearly separated from the remainder of the spectrum
by a large frequency gap. The thickness of the faces is crucial for
the stability of HK97, as a projection of the protein chains onto the
hypothetical icosahedral faces leads to the appearance of three
additional zero-modes. This yields a partial explanation for the
extremely low frequency at which the plateau appears for HK97. We
will return to the face thickness issue in section~\ref{s:faces}.

\begin{figure}[t]
\begin{center}
\includegraphics[width=.7\textwidth]{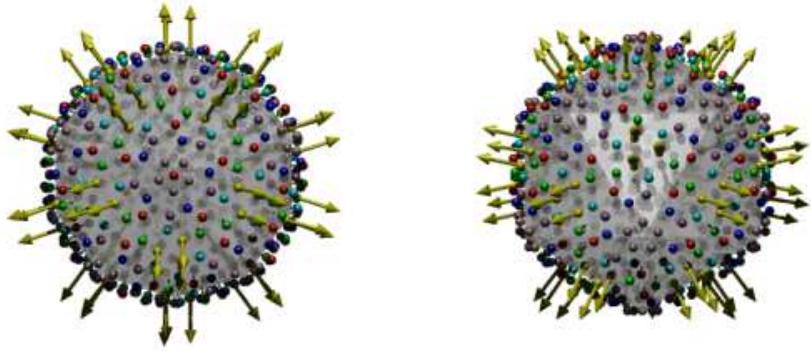}
\end{center}
\caption{The first singlet mode of HK97, viewed along a 5-fold
  symmetry axis (left) and along a 3-fold axis (right). It occurs at
  position 67 with our particular bond choices, which is arguably
  still within the regime of the low-frequency approximation. This
  mode exhibits protrusion of the hexamer units (or rather, one of the
  chains of the hexamer), as also found
  in~\cite{Kim:2003a}.\label{f:HK97_singlet}}
\end{figure}

The spectrum of HK97 has previously been analysed in the Tirion
approximation using RTB techniques by~\cite{Tama:2005a}. One of their
key results is that the first two conformal modes correspond to
protrusion of the hexamer and pentamer groups respectively. The
structure of these singlet modes of HK97 has also been analysed
by~\cite{Kim:2003a}. Our first singlet mode occurs well after the
plateau, and has a structure which confirms these more elaborate
computations. In particular, we see that the first singlet (depicted
in \figref{f:HK97_singlet}) exhibits the same type of hexamer
protrusion.

The structure of the 24-state low-frequency plateau which we observe
is in agreement with the analysis of~\cite{Rader:2005a}, who used an
all-atom Tirion potential and a Lanczos eigenvalue solver to find that
the first singlet occurs at position~31. However, we again note that
the use of more crude approximations, such as in
e.g.~\cite{Tama:2005a} or~\cite{Schuyler:2005a}, gives a low-frequency
structure that is different from ours.

\subsection{Simian virus 40}

\begin{figure}[t]
\begin{center}
\includegraphics[height=.3\textwidth]{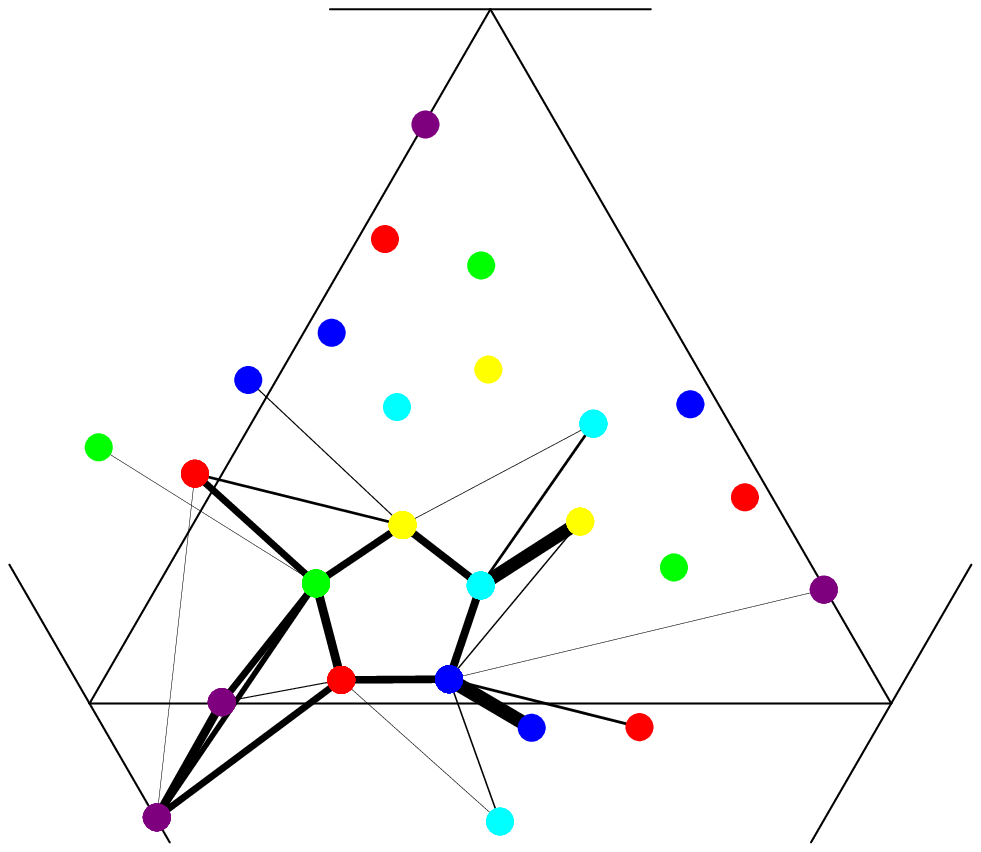}
\qquad
\includegraphics[height=.3\textwidth]{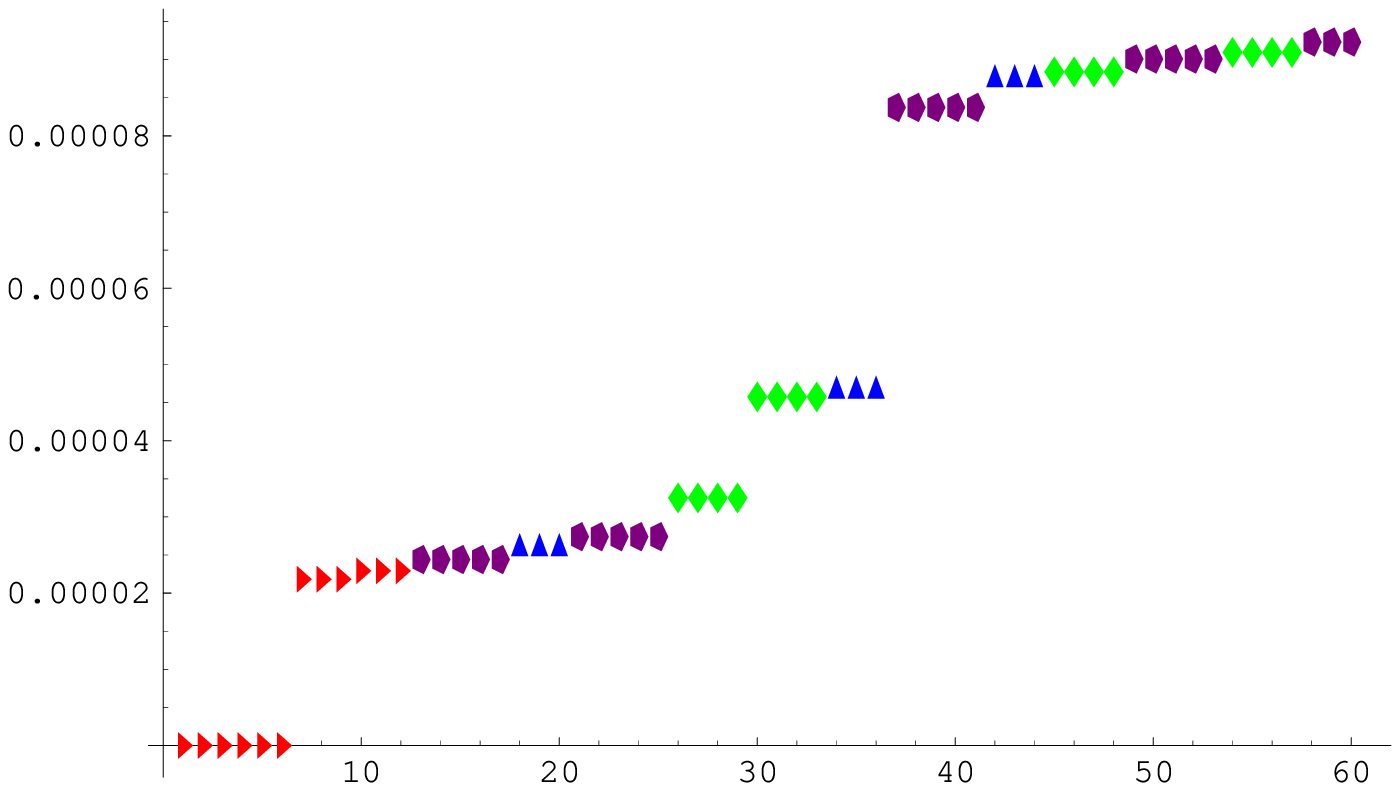}
\end{center}
\caption{The protein positions and inter-protein bonds for SV40 (left)
  as well as the low-frequency spectrum (right).\label{f:SV40_fund_freq}}
\end{figure}

Simian virus 40 has an all-pentamer capsid, which falls outside the
Caspar-Klug classification scheme. It is captured, however, by viral
tiling theory~\cite{Twarock:2004a}. This particularity set aside,
there is also a considerable difference in the bond structure as
compared to other viruses we have analysed here. While~HK97 exhibits
a rather rigid structure around the edge of the hexamer, SV40 has a
relatively weakly bound pentamer edge and much stronger bonds to other
pentamers. See the left panel of \figref{f:SV40_fund_freq} for
details.

The difference between SV40 and all other capsids discussed so far is
also clearly visible in the low-frequency spectrum: instead of the
24-state plateau which we have seen for all Caspar-Klug viruses, SV40
shows a much smoother structure. If a plateau is visible at all, it
now contains 30 states, with representation content
\begin{equation}
2\Gamma^{3}+2\Gamma^{3'}+2\Gamma^{4}+2\Gamma^{5}\,.
\end{equation}
We have so far not found any other virus with this low-frequency
structure in its spectrum, which suggests that it is not as universal
as the 24-state plateau observed for the Caspar-Klug family, but
confirming this will require a more elaborate scan through the
non-Caspar-Klug capsids.

\section{Generic patterns in frequency spectra of viral capsids}
\label{sec-generic patterns}
\subsection{The dodecahedron}
\label{s:dodecahedron}

In the previous section we presented the frequency spectra for a
selection of icosahedral viral capsids, calculated within the
mathematical set-up discussed in section~\ref{s:group theory}. We will
now argue that the distinctive low-frequency plateau of twenty-four
normal modes, observed for all Caspar-Klug capsids, is in fact a
direct manifestation of icosahedral symmetry. More precisely, we will
show that, after an appropriate separation of the degrees of freedom of
the viral capsids, an explanation of the low-frequency plateau can be
found in the vibrational spectrum of a simple dodecahedral spring-mass
model~\cite{Englert:2008a}.

In order to explain this, let us first focus on a caricature virus
capsid, namely an hypothetical capsid with only one protein chain per
icosahedral face, approximated by a point mass located at the centre
of the face. Such a system possesses $3\times 20 - 6 = 54$ degrees of
freedom after subtraction of the trivial rotation and translation
modes. We will also introduce a minimal set of bonds, obtained by
connecting the nearest-neighbour proteins, and view the latter as the
vertices of a polyhedron whose edges are the bonds. The resulting
structure has thirty edges and is a \emph{dodecahedron} dual to the
icosahedron naturally associated with the capsid considered.

\begin{figure}[t]
\begin{center}
\includegraphics[width=.45\textwidth]{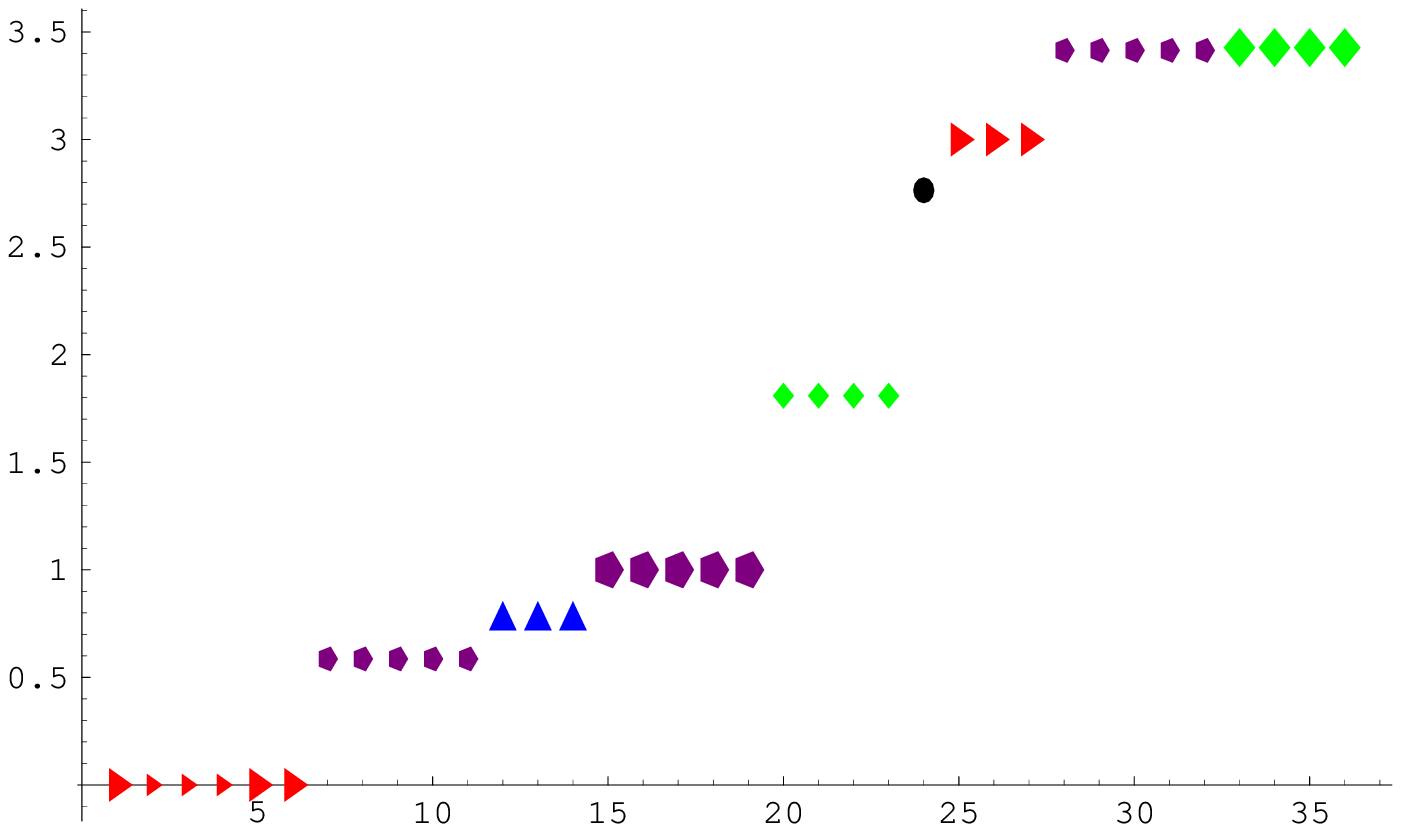}\qquad
\includegraphics[width=.45\textwidth]{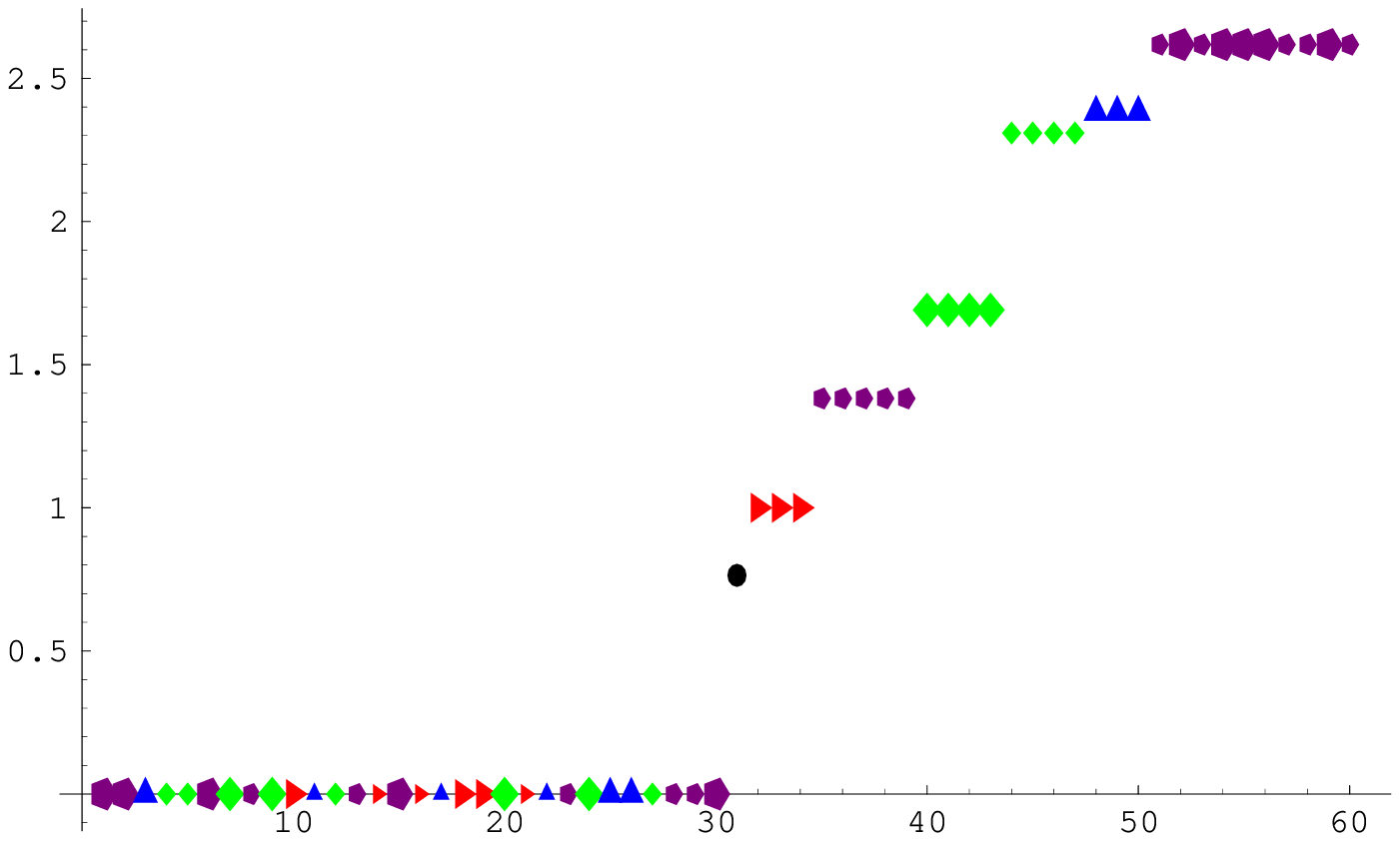}
\end{center}
\caption{The spectrum of the icosahedron (left) and dodecahedron
  (right), decomposed in irreducible representations of the
  icosahedral group. Large symbols denote ``$-$'' representations, small
  symbols denote ``$+$'' representations. Black dots denote singlets,
  red triangles~$\Gamma_{3^{\pm}}$, blue
  triangles~$\Gamma_{3'^{\pm}}$, green squares~$\Gamma_{4^{\pm}}$ and
  purple pentagons~$\Gamma_{5^{\pm}}$. Note that the structure of the
  non-zero modes is the same for both spring-mass models.\label{f:ico_dode}}
\end{figure}

The thirty bonds reduce the number of degrees of freedom from 54 to
24, and these remaining degrees of freedom correspond to the 24
zero-modes which signal the instabilities of the dodecahedral cage. We
now argue that the representation theory content of these zero-modes
coincides exactly with that of the low-frequency modes of Caspar-Klug
capsids. We first note that the dodecahedral spring-mass system
considered here possesses a centre of inversion. Therefore, the
relevant symmetry group is the full icosahedral group $H_3$ and the
decomposition of the displacement representation of the dodecahedral
capsid into a sum of irreducible representations of $H_3$ reads,
\begin{equation}\label{dode}
\Gamma^{\text{disp},60}_{\text{dode}} 
  = \Gamma^{1}_{+}\oplus \Gamma^{3}_{+}\oplus 2\Gamma^{3}_{-}\oplus 
    \Gamma^{3'}_{+}\oplus 2\Gamma^{3'}_{-}\oplus 
    2\Gamma^{4}_{+}\oplus 2\Gamma^{4}_{-}\oplus 3\Gamma^{5}_{+}\oplus
    2\Gamma^{5}_{-}\,.
\end{equation}
The numerical superscripts refer to the dimension of the
representations, while the $\pm$ signs differentiate between even and
odd representations. This decomposition does not make use of any
information stored in the force matrix, and therefore does not tell us
which of these modes are zero-modes. In order to pin those down, we
use the information encoded in the displacement representation of the
icosahedron to which the dodecahedron constructed above is dual. First
of all, the icosahedral vibration modes decompose as
\begin{equation}\label{ico}
\Gamma^{\text{disp},36}_{\text{ico}}
  = \Gamma^{1}_{+}\oplus \Gamma^{3}_{+}\oplus 2\Gamma^{3}_{-}\oplus 
    \Gamma^{3'}_{-}\oplus \Gamma^{4}_{+}\oplus \Gamma^{4}_{-}\oplus 
    2\Gamma^{5}_{+}\oplus \Gamma^{5}_{-}\,.
\end{equation}
In this context of vibrations, the link between the dodecahedron and
the icosahedron comes from considering the subset of motions of the
dodecahedral system which are induced by the motion of the 12 vertices
of an icosahedron. By ``induced'' we mean that the dodecahedron moves
in such a way that its vertices are located at the centre of the
deformed icosahedral faces, at all times. All icosahedral modes
belonging to a particular irreducible representation must induce a
linear combination of modes of the dodecahedral capsid which pertain
to the same irreducible representation. Hence, provided that the
vibrational modes of the icosahedron induce a non-vanishing component
in the finite-frequency modes of the dodecahedron, we can conclude
that the modes in~$\Gamma^{\text{disp},60}_{\text{dode}}$ which are
not contained in~$\Gamma^{\text{disp},36}_{\text{ico}}$ must have
vanishing frequency. Under this assumption, the zero-modes of the
dodecahedron transform as
\begin{equation}
\Gamma^{\text{disp}}_{\text{dode, zero}} = 
   \Gamma^{3'}_{+}+\Gamma^{3'}_{-}+\Gamma^{4}_{+} + \Gamma^{4}_{-} + \Gamma^{5}_{+}+\Gamma^{5}_{-}\,.
\end{equation}
As a matter of fact, an explicit calculation of the normal modes confirms this
assumption. The full frequency spectrum of the icosahedron and dodecahedron is
displayed in~\figref{f:ico_dode} for reference.

\begin{figure}
\begin{center}
\includegraphics[width=.7\textwidth]{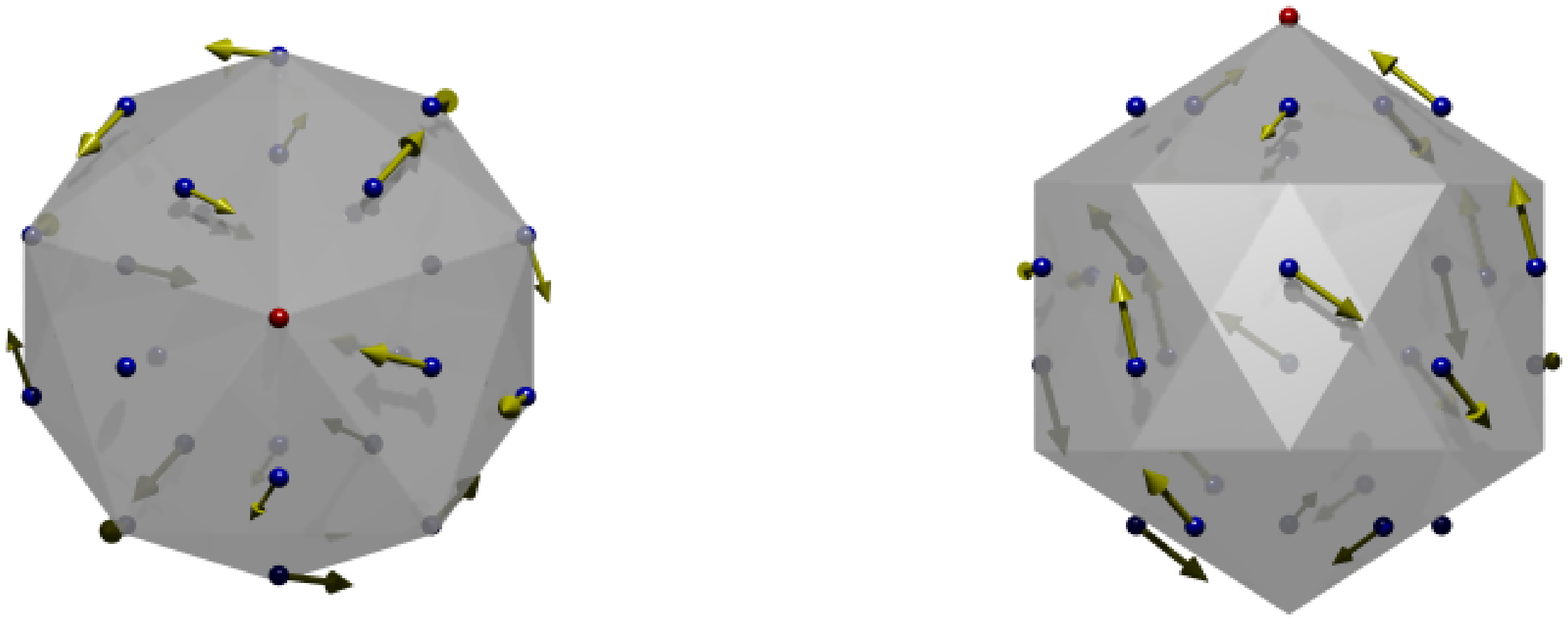}\vspace{-4ex}
\includegraphics[width=.7\textwidth]{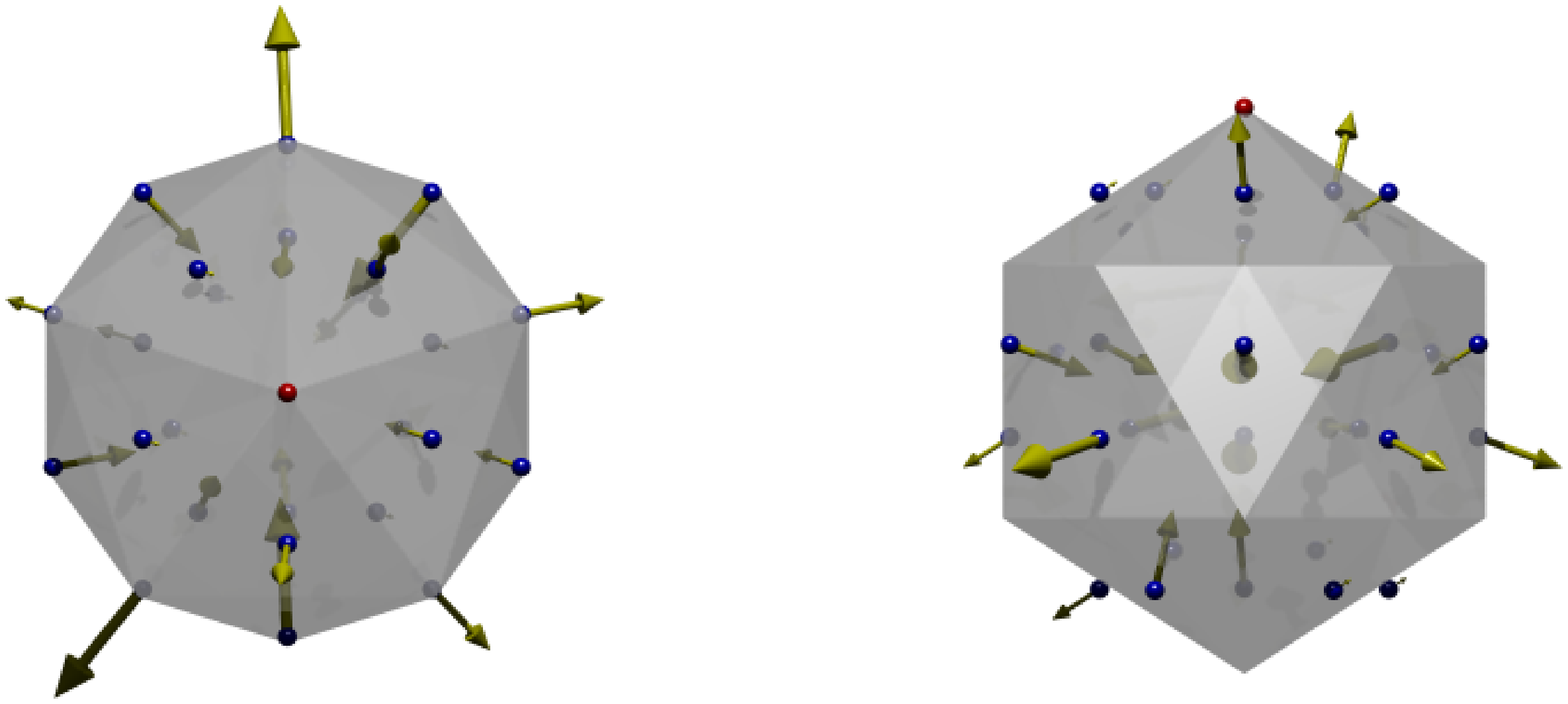}
\vspace{-6ex}
\end{center}
\caption{The zero-modes of the dodecahedron in the $\Gamma_{3'^+}$
  (top) and $\Gamma_{3'^-}$ (bottom)  representations.\label{f:dode3prime}}
\end{figure}

\begin{figure}
\begin{center}
\includegraphics[width=.7\textwidth]{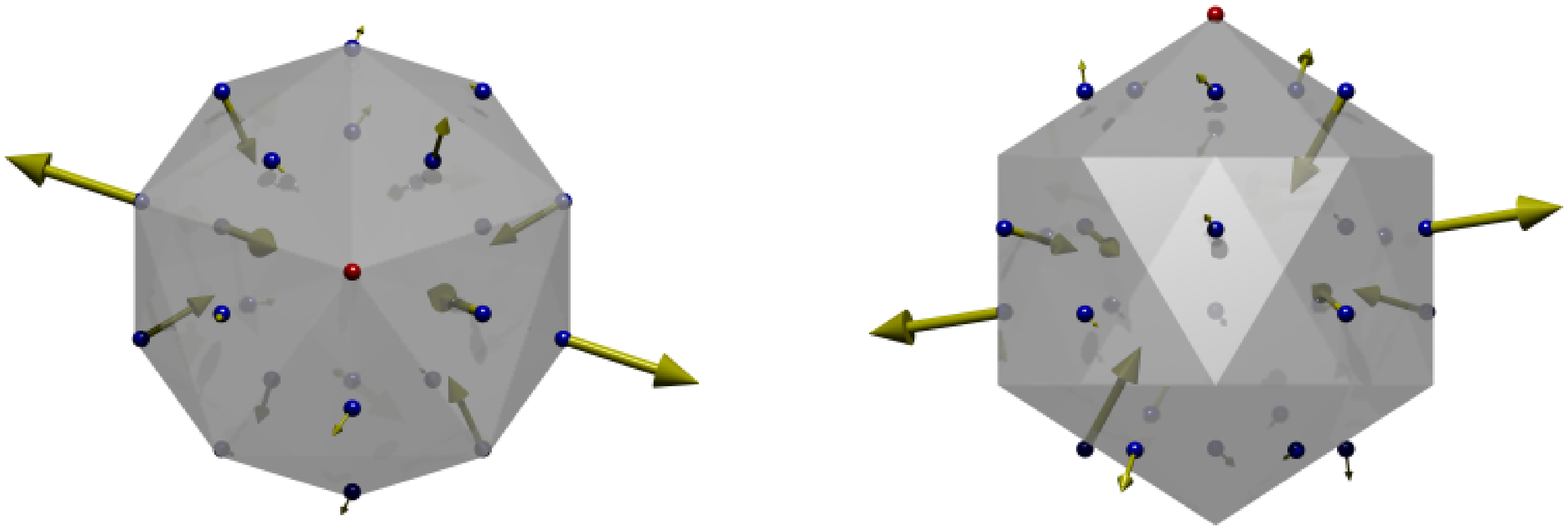}\vspace{-4ex}
\includegraphics[width=.7\textwidth]{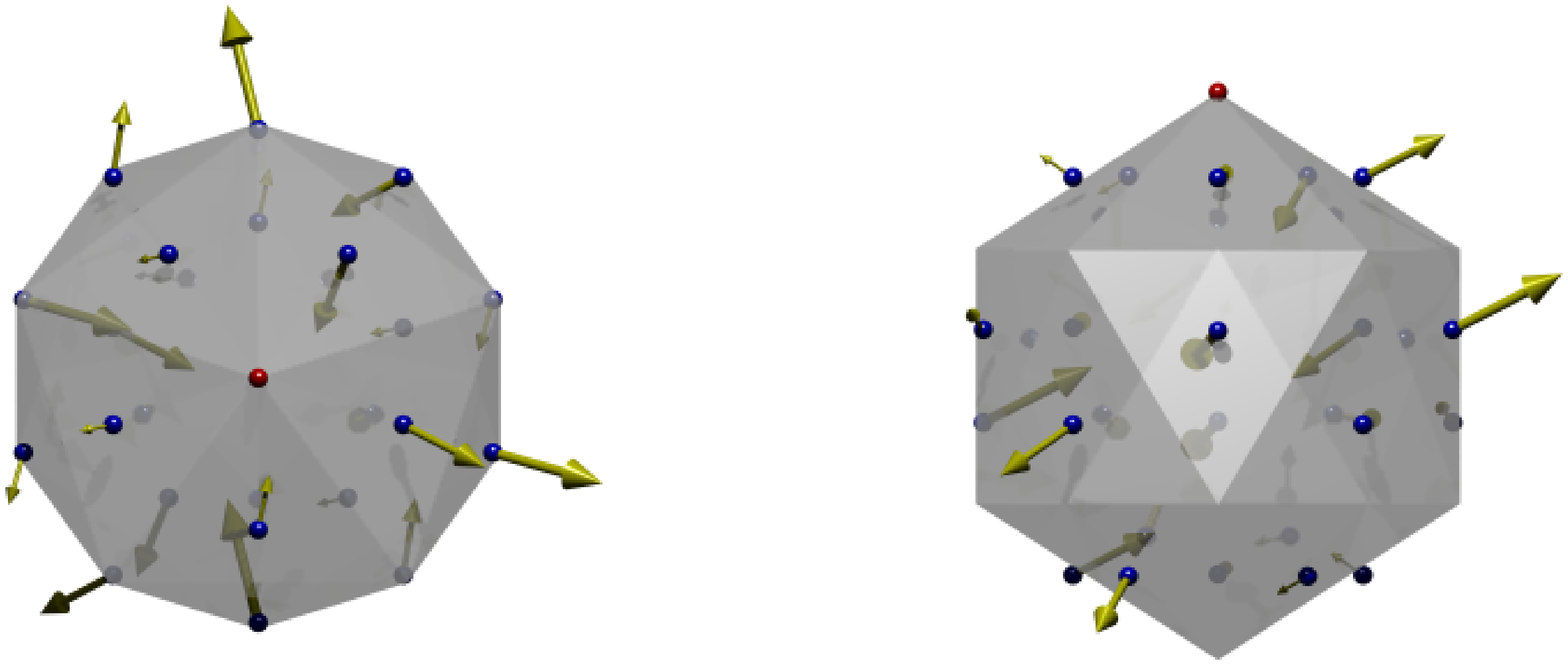}
\vspace{-6ex}
\end{center}
\caption{The zero-modes of the dodecahedron in the $\Gamma_{4^+}$ (top)
  and $\Gamma_{4^-}$ (bottom) representations.\label{f:dode4}}
\end{figure}

\begin{figure}
\begin{center}
\includegraphics[width=.7\textwidth]{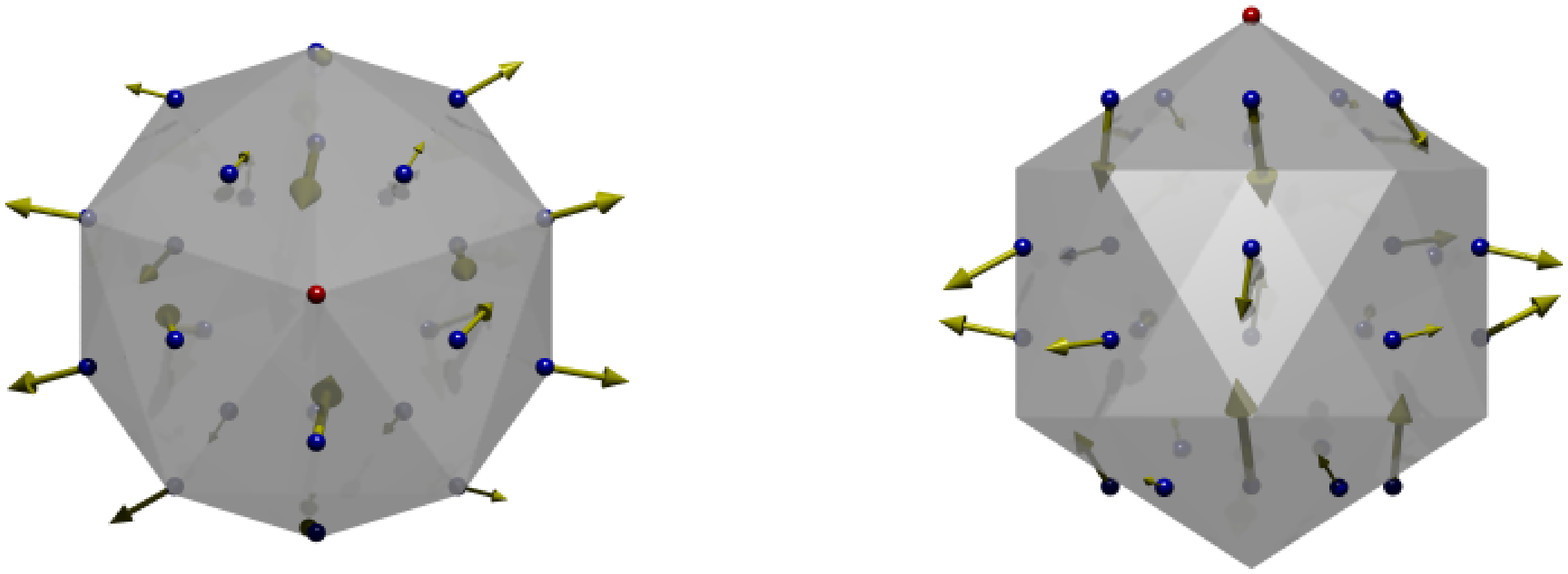}\vspace{-4ex}
\includegraphics[width=.7\textwidth]{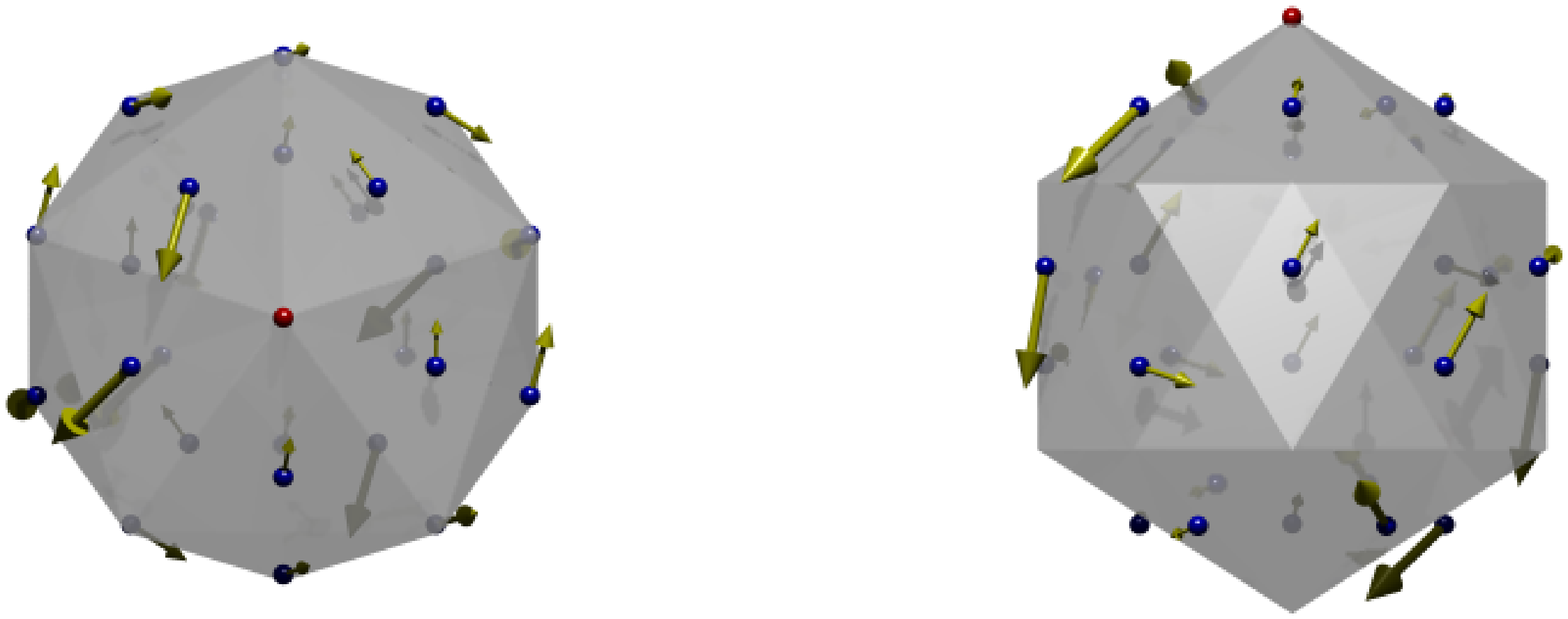}
\vspace{-6ex}
\end{center}
\caption{The zero-modes of the dodecahedron in the $\Gamma_{5^+}$ (top)
  and $\Gamma_{5^-}$ (bottom) representation.\label{f:dode5}}
\end{figure}

We have visualised the twenty-four zero-modes in
\figref{f:dode3prime}--\ref{f:dode5} and use the graphical
representations to compare with the low-frequency modes of actual
Caspar-Klug viruses.\footnote{Animations of the dodecahedron vibration
  modes are available at \url{http://biomaths.org/}.} In comparing
with those of the Polio capsid in~\cite{Vlijmen:2005a}, one should
recall that we perform a linearised analysis, and thus, any linear
combination of eigenvectors with the same eigenvalue remains an
eigenvector.

We end up this subsection with some insights on the multiplicity of
the 24 zero-modes. Let us start with the 5-fold degenerate
representations.  Motion in the~$\Gamma_{5^+}$ representation is
associated to squeezing in the direction of one of the six 5-fold axes
of the icosahedron. A linear combination of all of them produces an
icosahedrically symmetric motion, which is a singlet, so the
degeneracy of the~$\Gamma_{5^+}$ modes is indeed five. Similarly, the
modes in the $\Gamma_{5^-}$ representation correspond to motion in
which two opposite caps centred on a five-fold axis are rotated in
opposite directions. Again, the linear sum produces a singlet, so the
degeneracy is 5-fold. The~$\Gamma_{4^+}$ and~$\Gamma_{4^-}$ modes are
related to the five cubes which are inscribed in the dodecahedral
cage; again one linear combination transforms as a singlet (see
also~\cite{Baskerville:1999ve}). Generically, the dimensions of the
irreducible representations are thus one lower than the multiplicity
of modes with a similar geometrical pattern.

\subsection{Faces as building blocks}
\label{s:faces}

The analysis of the vibrations of the caricature virus capsid has
revealed the existence of 24 zero-frequency modes. Despite its
simplicity, this model can nevertheless be used to understand the
low-frequency pattern of more complicated capsids, such as those
analysed in section~\ref{s:alltogether}. The key ingredient here is to
first analyse the vibration modes of the proteins on each icosahedral
face (for the caricature capsid above, those vibration modes are just
the three trivial translation modes).

\begin{figure}[th]
\begin{center}
\includegraphics[width=.4\textwidth]{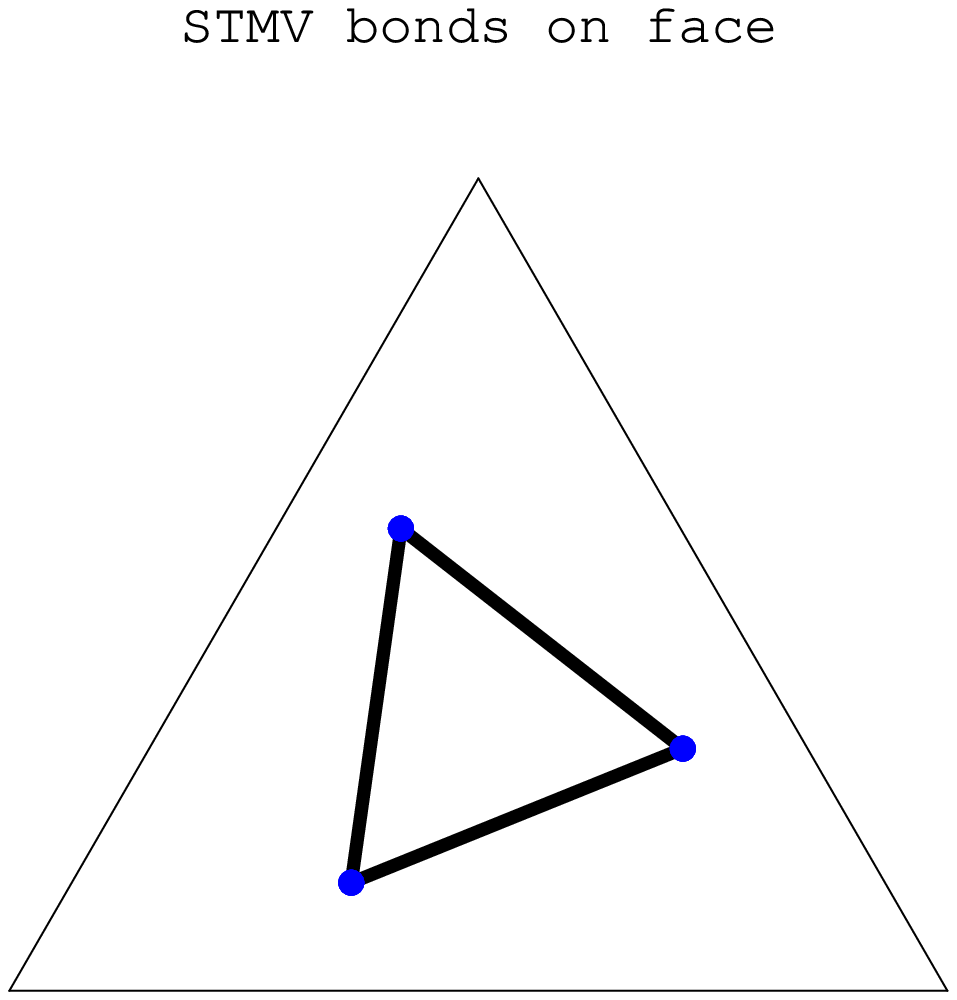}
\includegraphics[width=.4\textwidth]{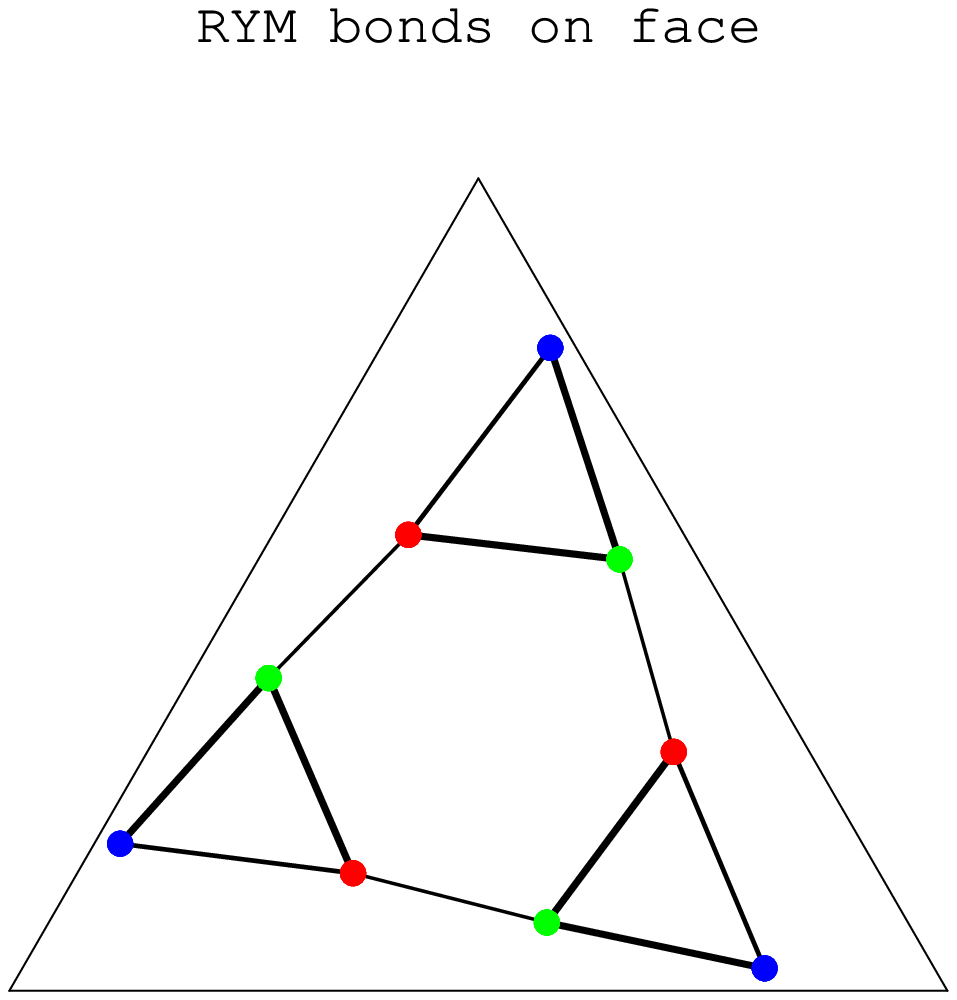}
\vspace{4ex}

\includegraphics[width=.4\textwidth]{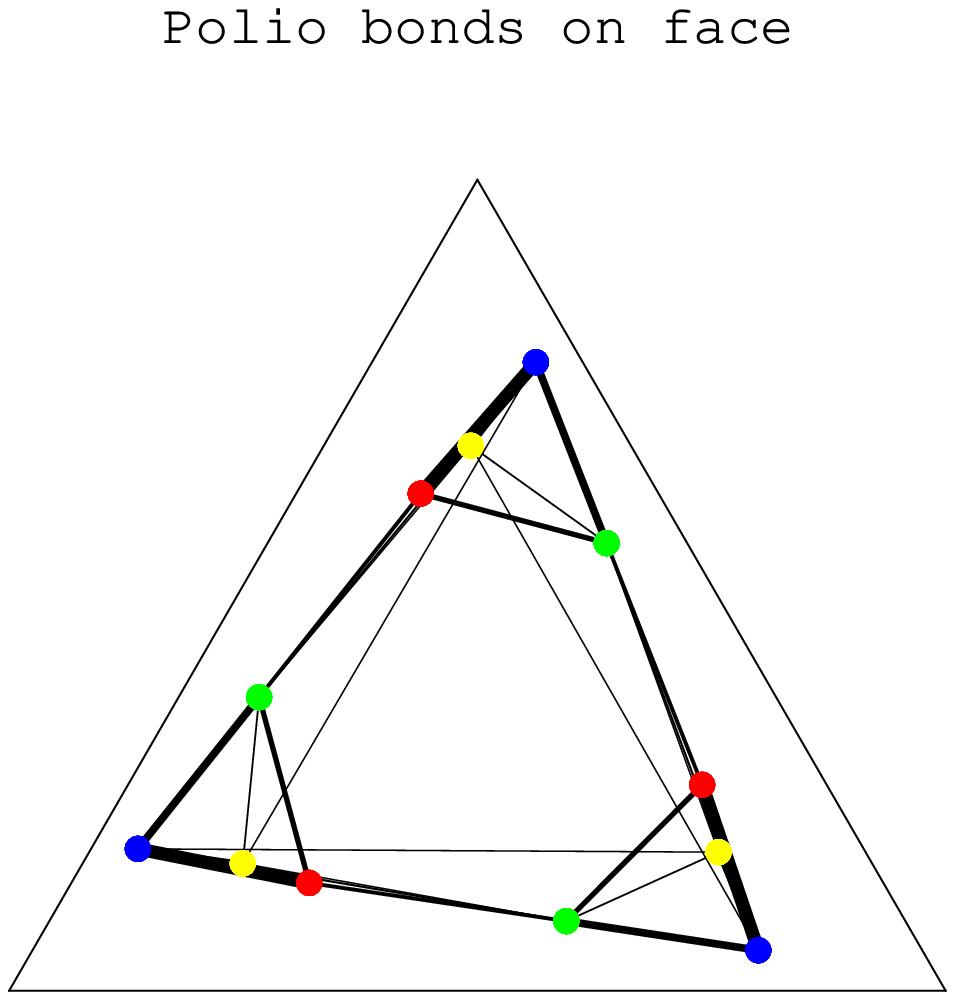}
\includegraphics[width=.4\textwidth]{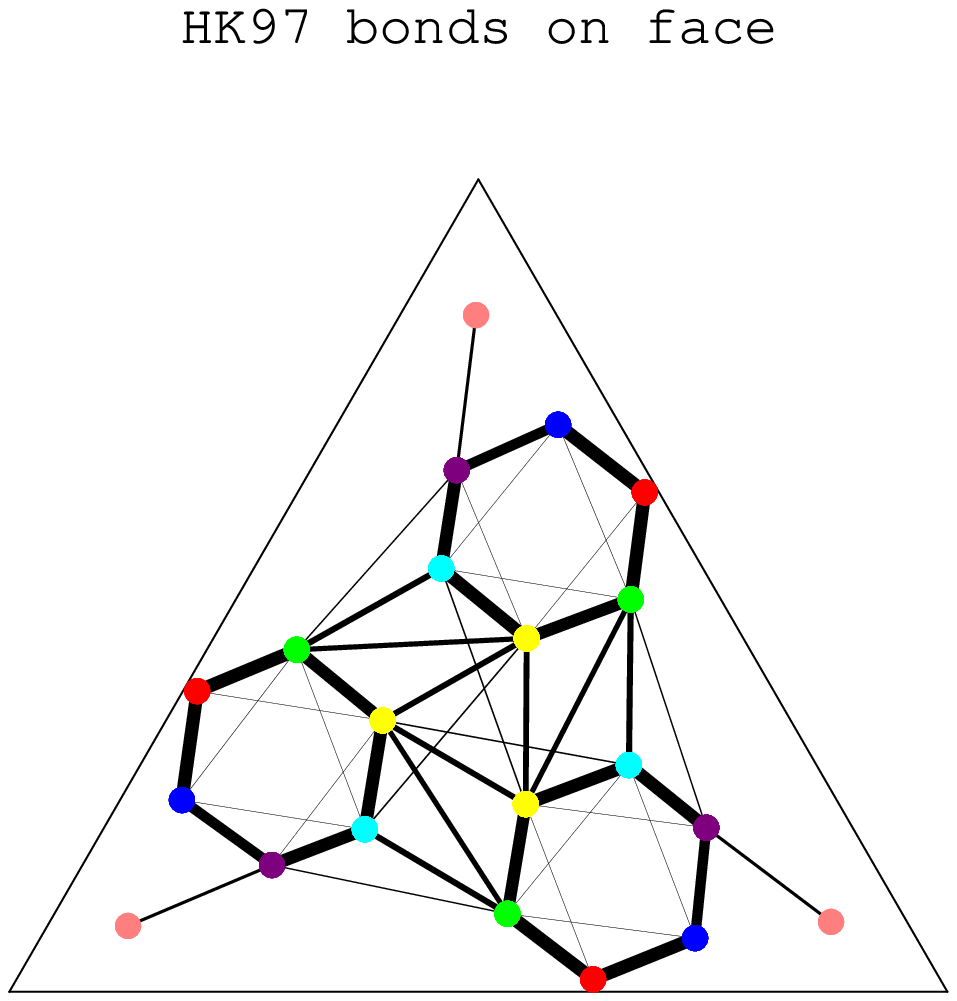}
\vspace{-1ex}
\end{center}
\caption{Structure of bonds on the faces of various virus
  capsids. These faces have 6(6), 15(15), 6(15) and 12(27) zero-modes
  respectively, where numbers in brackets include the infinitesimal
  instabilities which go away once the ``thickness'' of the face is
  taken into account.\label{f:face_bonds}}
\end{figure}

\begin{figure}[th]
\begin{center}
\includegraphics[width=.4\textwidth]{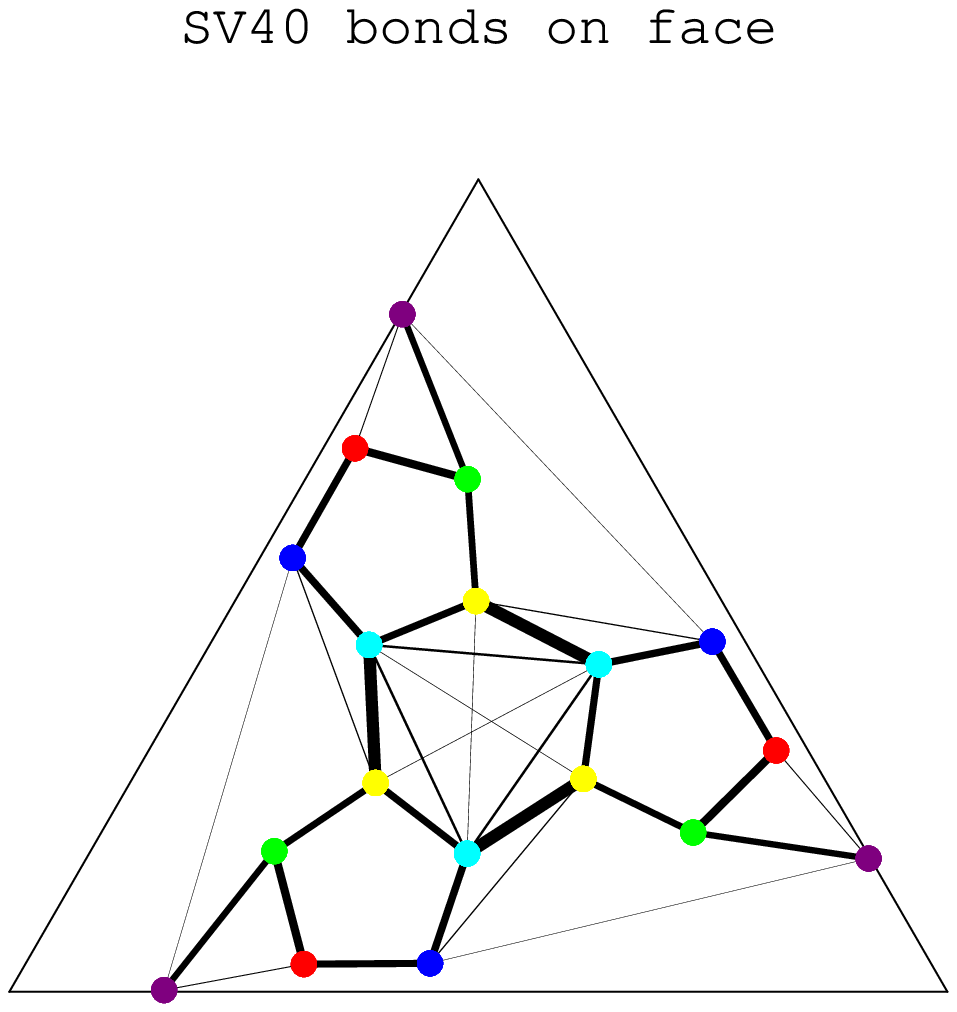}
\end{center}
\caption{Structure of bonds on the face of SV40. Its faces have 18(21)
  zero-modes respectively (notation as in
  figure~\ref{f:face_bonds}).\label{f:face_bonds2}}
\end{figure}

Before we explain this logic in detail, let us summarise the situation
for the various capsids which we have discussed in
section~\ref{s:alltogether}.  In \figref{f:face_bonds}
(resp.~\ref{f:face_bonds2}) we display the non-zero bonds of STMV, RYMV,
Polio, HK97 (resp.~SV40), restricted to those bonds which
remain on a single icosahedral face. A crucial observation is that the
faces are rarely stable: only in the case of STMV is there a
sufficient number of inter-protein bonds to prevent the face from
collapsing. The faces themselves thus exhibit zero-modes in their
eigenspectrum. Additional bonds, which reach between proteins on
different faces, are necessary for stability of the capsid as a whole
and the faces in particular.

\begin{figure}[th]
\begin{center}
\vspace{3ex}
\includegraphics[width=.8\textwidth]{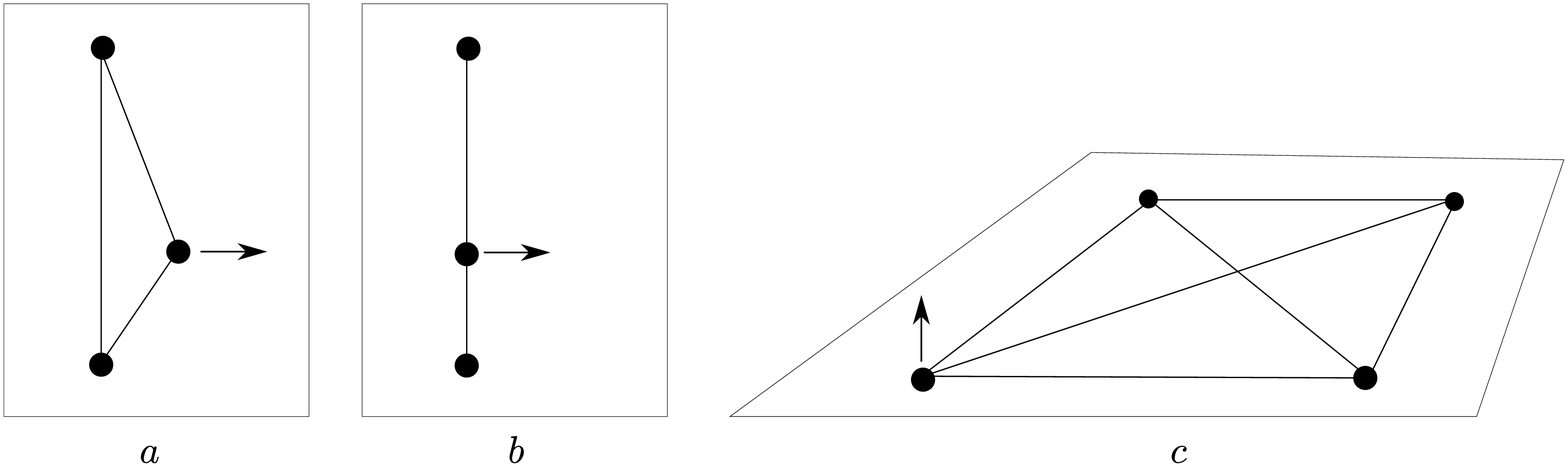}
\vspace{-3ex}
\end{center}
\caption{Infinitesimal instabilities arise when the restoring force
  vanishes at linear order. The 2d structure in figure~$a$ is stable,
  while the structures in~$b$ and~$c$ exhibit an infinitesimal
  instability.\label{f:infinitesimal_instability}}
\end{figure}

It is interesting to comment at this point on how zero modes arise. If
all proteins associated with a given icosahedral face do actually lie
in the plane of that face, ``infinitesimal instabilities'' may
develop, as the forces that can be exerted by all bonds connected to a
node lie in a plane. The force in the direction normal to the plane
then vanishes at linear order, leading to a zero-mode of the force
matrix. Figure~\ref{f:infinitesimal_instability} illustrates these
infinitesimal instabilities in two and three dimensions.  Generically,
a fully connected network in the plane with $n$~nodes has
$n$~zero-modes which move the individual nodes out of the plane (the
$z$~direction). Of these, one is a translation mode, and two others
are rotations around the $x$- and $y$-axis respectively. Therefore,
there are $n-3$ modes which are associated with infinitesimal
instability.
In the real world, the faces of the icosahedral capsids always have a
finite thickness, because the distance of the protein chains to the
centre of the capsid is not the same for all chains. As soon as this
thickness is introduced, all infinitesimal instabilities
disappear. Nevertheless, due to the fact that the faces are relatively
thin, these lifted zero-modes will often have a small frequency.

Let us now turn to the general argument.  A $T=n$ Caspar-Klug capsid has~$3n$
proteins per face, and we will denote the number of links between
these proteins with~$3k$. If there were no links between proteins
on different faces, one would expect a total of
\begin{equation}
N_0 = (9n - 3k)\cdot 20 = 60 (3n -k)
\end{equation}
capsid zero-modes. A stable capsid, which only has 3 rotational and 3
translational zero-modes, would thus require at least $60(3n-k)-6$
independent constraints (i.e.~bonds). How these bonds should be chosen
remains an open mathematical problem. In two dimensions there exists
an elegant algorithm to determine the stability of a network. This
``pebble game'' algorithm scales approximately linearly in the number
of nodes~\cite{Jacobs:1995a}. The crucial ingredient which makes this
possible is the Laman theorem~\cite{Laman:1970a}, which allows one to
find redundant bonds by analysing strict subgraphs of the full
network. This theorem does not generalise to higher dimensions, and
despite considerable effort an exact analogue of the pebble game in
three dimensions is not known.

The VIPERdb database, however, provides us with various association
energies for edge-crossing bonds which we can use to stabilise the
capsids. By symmetry, these bonds come in multiples of 30 or 60 when
one considers the capsid as a whole. The logic above suggests that, if
we keep the strongest $e = 2(3n-k)-1$ bonds per edge, this would yield
a capsid with 24 non-trivial zero-modes. A careful analysis of the
``on-face'' bond structure given earlier confirms that this indeed is
the case for all Caspar-Klug capsids which we have analysed. Here are some examples.

\begin{figure}[t]
\vspace{-2ex}
\begin{center}
\includegraphics[width=\textwidth]{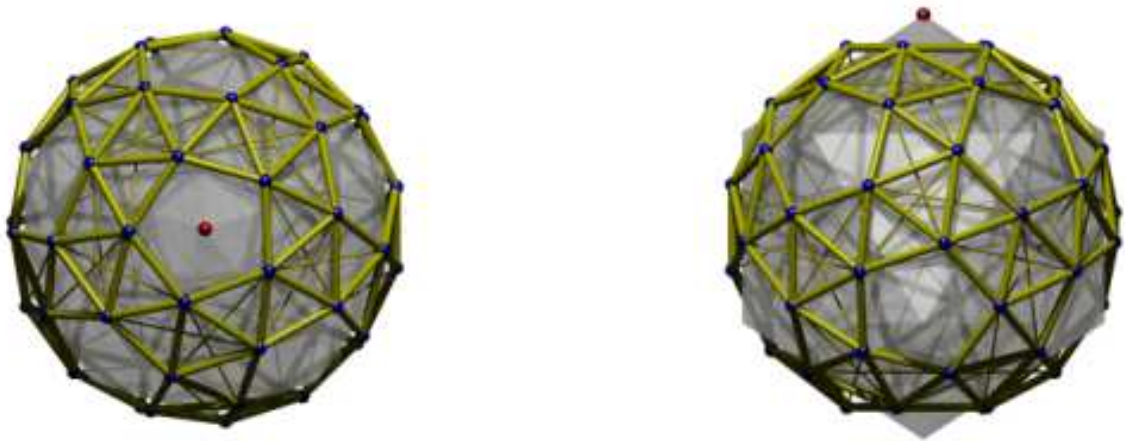}
\end{center}
\vspace{-6ex}
\caption{``Top'' and ``front'' view of the capsid of STMV, including
  the inter-protein bonds. Dots denote the centre-of-mass positions of
  the protein chains, tubes denote bonds between them. This structure
  has 30 zero-modes without the A1-A8 bond (thin tubes); 24 of these
  are lifted by this weak bond.\label{f:STMV3d}}
\end{figure}

The STMV capsid, with~$n=1$, is the simplest case to analyse. It
has a stable triangular face structure with~$k=1$, and from
$2(3n-k)-1=3$ we expect to need $e=3$ edge-crossing bonds per edge.
Looking at the left panel of \figref{f:STMV_fund_freq} we observe that
if we remove the weakest (longest) bond from the capsid, there are
indeed 3 bonds per edge (the figure shows two of these bonds, of which
one does not go through a two-fold axis and therefore really appears
twice per edge after the icosahedral symmetries are taken into
account). This reduced capsid is visualised in \figref{f:STMV3d}. An
analysis of its spectrum indeed exhibits 24 non-trivial zero-modes in
addition to the 6 translation and rotation modes, in agreement with
the counting argument given above.

For the RYMV capsid, which has~$n=3$, if we discard the two weakest
bonds (which are both edge-crossing) we end up with~$k=4$ on-face and
$e=9$ edge-crossing bonds, and 24 non-trivial zero-modes, in
accordance with~$2(3n-k)-1=9$. As discussed in
section~\ref{s:RYM_TBSV_CCMV}, the bonds that stretch from the
C-chains are crucial for stability, and their weakness is responsible
for the fact that the zero-modes are only slightly lifted. For HK97,
which has~$n=7$, the situation is more subtle. Some of the
weakest bonds on the faces of this capsid are actually weaker than
some of the edge-crossing bonds. Consistently removing the weakest six
bonds leads to a capsid with~$k=12$ and $e=17$. The expression
with~$2(3n-k)-1=17$ thus suggests that in this case there should be 24
non-trivial zero-modes, which we indeed observe. This type of analysis
can be extended to all the other Caspar-Klug viruses, with similar
results.



%



\subsection{Tiling independence}

We have seen that, despite the considerable difference in the
structure of Caspar-Klug capsids, the low-frequency spectrum exhibits
a remarkable uniformity. A priori one might have expected that
the way in which the strongest bonds are distributed over the capsid
has a determining influence on the low-frequency vibration pattern. 

\begin{figure}[t]
\begin{center}
\includegraphics[width=.32\textwidth]{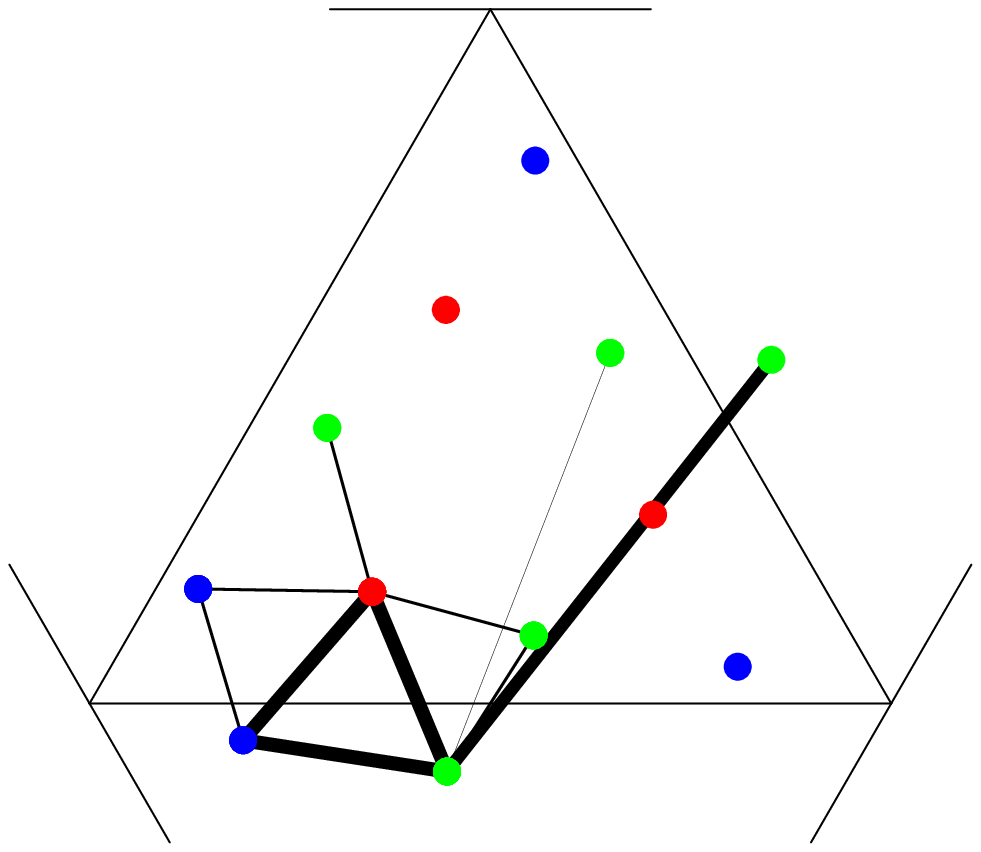}
\includegraphics[width=.32\textwidth]{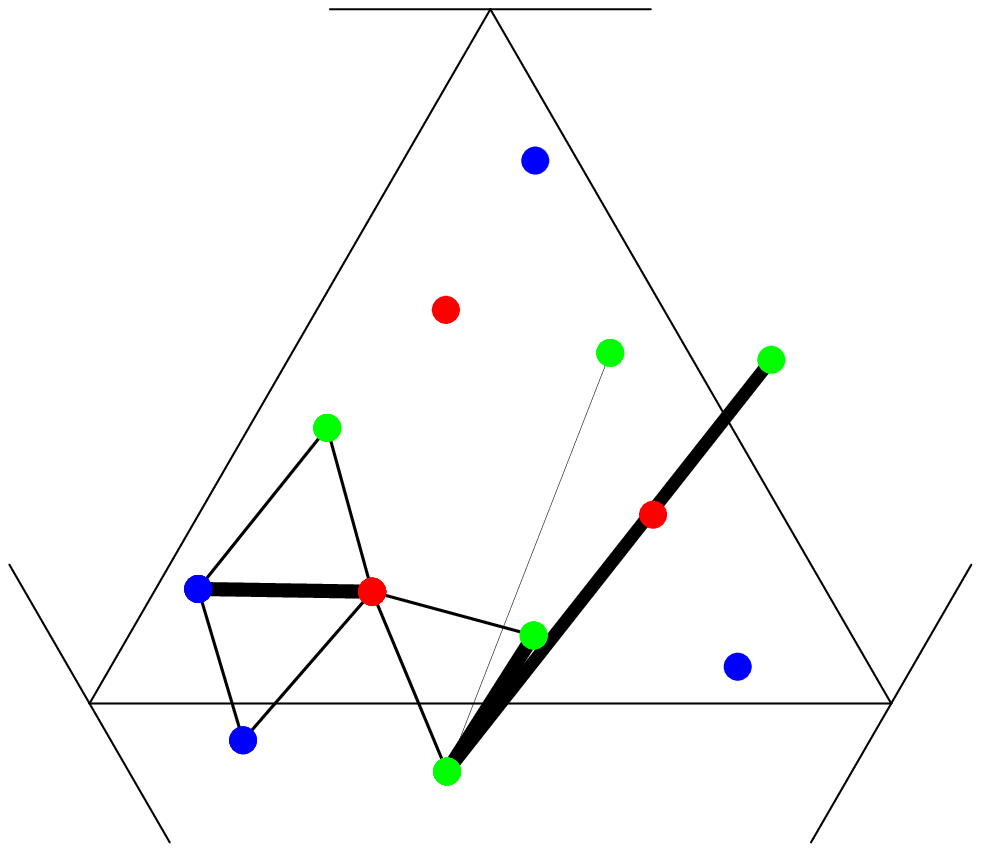}
\includegraphics[width=.32\textwidth]{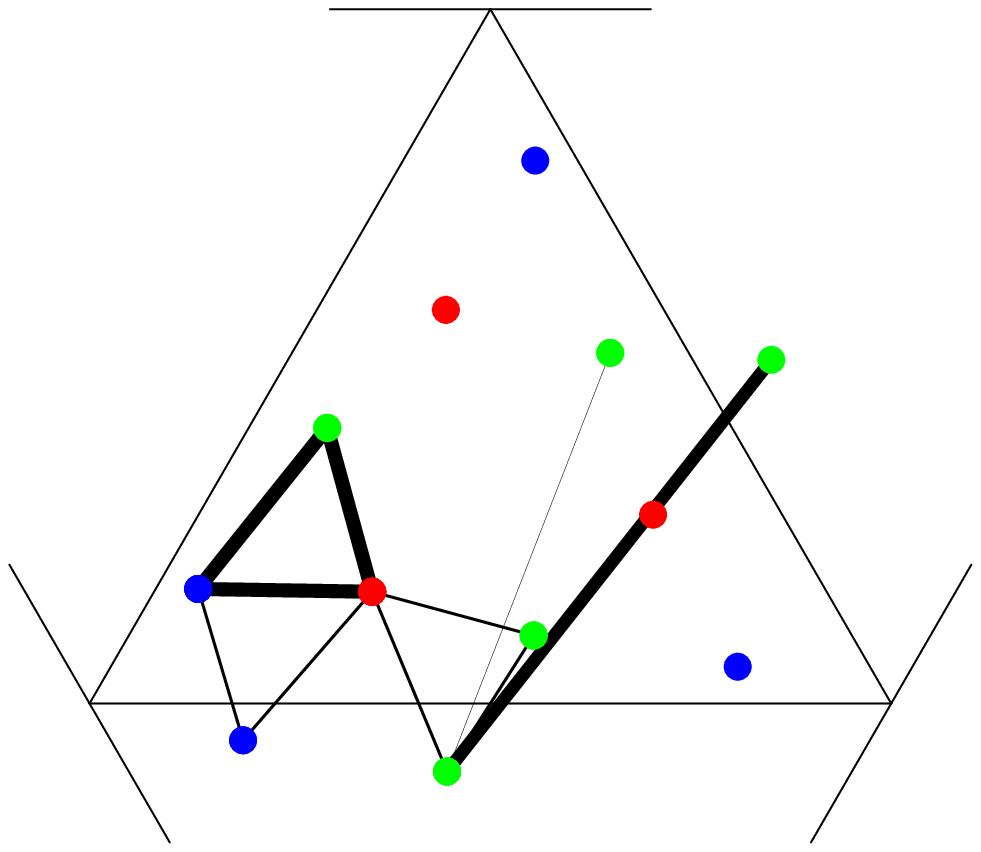}
\caption{Three hypothetical virus capsids, corresponding to a
  triangle, rhomb and kite bond structure
  respectively.\label{f:hypo_capsids}}
\end{center}
\vspace{2ex}
\begin{center}
\includegraphics[width=.95\textwidth]{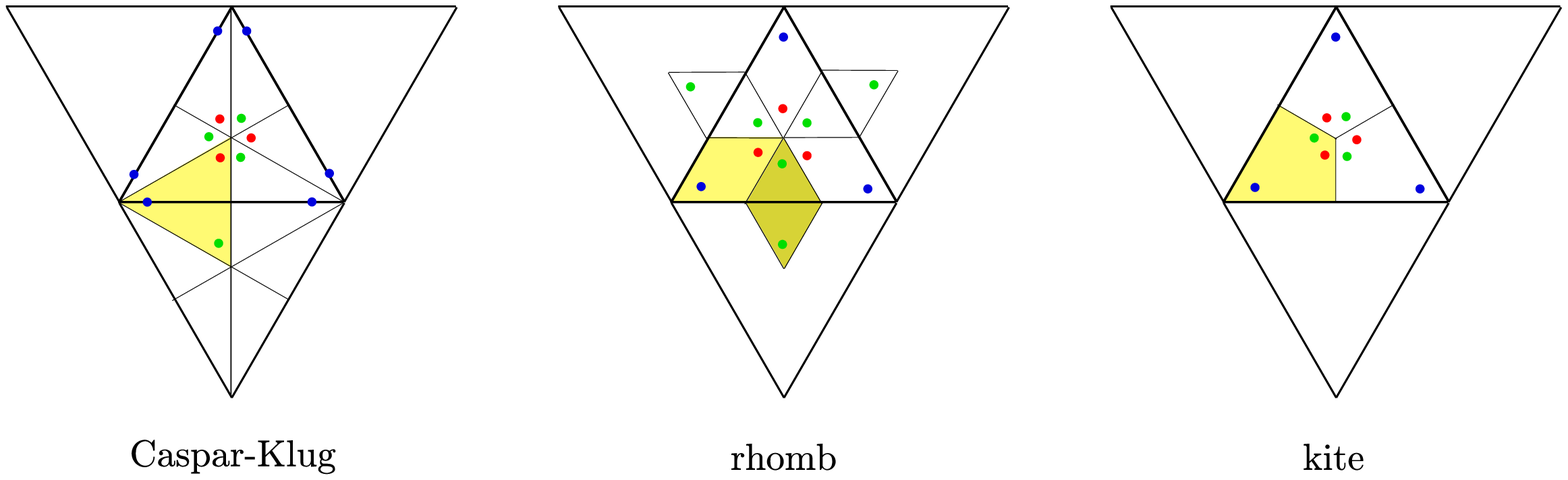}
\end{center}
\vspace{-2ex}
\caption{The three possible tilings for~$T=3$ capsids. The
  central triangle denotes one face of the icosahedron. The strongest
  inter-protein bonds are those between proteins on a single tile,
  i.e.~on one of the yellow-coloured polygons. Bonds between
  equal-type chains are all \emph{weak} in the Caspar-Klug and kite
  tilings, while some of them are \emph{strong} in the rhomb
  tiling. The rhomb tiling is the only one without strong bonds
  between monomers on different faces.\label{f:t3tilings}}
\end{figure}

In order to isolate the effects of variations in the bonding patterns
from the effects of changing protein locations, we have analysed three
hypothetical $T=3$ capsids. These are all based on the protein
positions of RYMV, but have different bond types corresponding to the
three different $T=3$ tilings: triangle, rhomb or
kite. In \figref{f:hypo_capsids} we display these hypothetical
capsids, while in~\ref{f:t3tilings} we show the three possible
tilings in their most ideal form. One can think of the capsids in
\figref{f:hypo_capsids} as small deformations of RYMV, CCMV and
Polio (without 4th chain), respectively, as their protein positions
are very similar. We have taken a hierarchy of bond strengths in which
the strongest bonds are of unit strength, most other bonds are of
strength~$1/5$, and the long C-chain arms are taken to be of
strength~$\alpha_1$ and $\alpha_2$ respectively. The positions of the
bonds are based on those of RYMV, ignoring the weak long-range bond
between the B-~and C-chain.

An analysis of the low-frequency spectrum exhibits a 24-state plateau
in all cases, provided that the bonds~$\alpha_1$ and~$\alpha_2$ are
not simultaneously taken to be very small. Thus, the size of the gap
is at least partially caused by one of the long-range bonds not being
extremely small. When one of the~$\alpha_{1,2}$ is set to zero, the 24
low-frequency modes come down to zero frequency. Interestingly, when
both long-range bonds are very small, i.e.~$\alpha_1=\alpha_2\ll 1$,
the plateau extends to a total length of 64. In all three capsids, one
then finds a state at position 71 which is a singlet conformal mode.

We thus conclude that the structure of the bonds is far less relevant
than one would expect, and the spectrum is essentially determined by
the few long-range edge-crossing bonds which make the capsid
stable. This provides some support for the analysis in
section~\ref{s:dodecahedron}, in which most results were obtained by
analysing the displacement representation, which contains no
information about bonds. However, there clearly are many mathematical
questions which these observations raise, and we hope to return to
some of these in a future publication.

\section{Discussion and conclusions}

Dynamical properties of viral capsids are undoubtedly key to our
understanding of conformational changes, maturation and function of
viruses. They may even offer clues on how to interfere mechanically
and chemically with their assembly or, more generally, their
replication cycle, with potentially important repercussions in the
health sector.

In this paper, we have been concerned with slow, large amplitude
vibrations of icosahedral capsids, which are thought to be crucial for
the onset of conformational changes. The underlying techniques, which
have been used for decades in the analysis of single protein
vibrations, are in principle applicable in the present context, but
the complexity of large biomolecular assemblies renders molecular
dynamics simulations prohibitive in cost and computer time. A major
tool in the study of viral capsid vibrations is Normal Mode Analysis
(NMA), despite its limitations, briefly mentioned in
section~\ref{s:overview}. The basic inputs of a NMA are the chosen
level of coarse-graining for the system under consideration and the
quadratic potential adopted. The few existing analyses are probing
different aspects of virus vibrations, and are just too scarce to
provide good ground for identifying general patterns across viruses.

Our approach has been to study low-frequency modes of vibration from a
coarse-grained approximation which replaces each capsid protein by a
point mass located at its centre of mass, while interactions are
dictated by a spring-mass model with all masses being equal and spring
constants reflecting the relative values of association energies
listed in VIPERdb.  Our analysis further assumes that the capsid is
empty.  Although our coarse-graining is certainly drastic, its
combination with a quadratic potential sensitive to a variety of bond
strengths has enabled us to detect universal properties of the
frequency spectra, namely a low-frequency plateau of twenty-four
states, which is present in all Caspar-Klug capsids analysed, i.e.~in
all capsids exhibiting twelve pentamers at the vertices of an
icosahedron and hexamers respecting the icosahedral symmetry
everywhere else, irrespectively of the tiling type (triangles, rhombs
or kites). Such signature is difficult to extract from most available
molecular dynamics simulations we are aware of. In some cases
(e.g.~\cite{Tama:2002a,Tama:2005a}), we believe the reason is that the
distance cut-off needed to stabilise capsids described by a Tirion
potential smears the low-frequency spectrum and loses the plateau
structures, a phenomenon illustrated in a toy model in
appendix~\ref{app-Tirion}.  Another potential source of confusion
comes from the fact some analyses only keep normal modes of vibration
which are invariant under the proper rotation subgroup ${\cal I}$ of
the icosahedral group, while we argued in~\ref{s:symmetry-based} that
although the potential must be invariant, the solutions to the
equations of motion need not be. Yet, there is one analysis which is
compatible with our results: in the all-atom simulation with cut-off
7.3\AA~of the HK97 capsid in~\cite{Rader:2005a}, the first
non-degenerate mode (a singlet of ${\cal I}$) appears at position 31
in the spectrum, in total agreement with our qualitative arguments on
the number of low-frequency modes (6 trivial zero-modes and 24 very
near zero-modes).

Our top-down approach is therefore complementary to the all-atom
simulations, and paves the way for a new generation of models which
should address several shortcomings of the present methods. Within the
harmonic approximation, a very interesting and deep mathematical
question is how to minimise the number of bonds in a 3-dimensional
spring-mass model so that the structure remains stable. In the absence
of useful theorems, one can be guided by Nature and make use of the
data on association energies in VIPERdb. However, there are several
viral capsids for which interesting dynamics has been observed using
the RTB method or all-atom normal mode computations, but which cannot
currently be handled with the coarse-grained model proposed in this
paper. This is due to the limited number of bonds listed in
VIPERdb. Capsids for which the available bonds are insufficient, even
after introduction by hand of one or two additional bonds, include
Hepatitis~B and MS2. In this light, it would be interesting to revisit
the computation of the association energies for these capsids. Another
outstanding puzzle is a proper mathematical understanding of the
different low-frequency plateau structure of SV40, in particular
whether the 30 near zero-modes are a signature of all-pentamer
capsids.

Finally, in view of the importance of conformational changes in viral
capsids, which can only be fully characterised by going beyond the
harmonic approximation, it is desirable to develop methods taking
anharmonicity into account. For instance, one might consider
generalising to viral capsids the Elastic Network Interpolation method
developed in~\cite{Kim:2006a} to generate anharmonic pathways of
macro-biomolecules.

\vfill
\section*{Acknowledgements}

We thank Fran\c{c}ois Englert for an inspiring collaboration on some
parts of the work reported here (see \cite{Englert:2008a}). We also
thank Peter Stockley, Roman Tuma and especially Reidun Twarock for discussions.

\newpage
\appendix
\renewcommand{\theequation}{\Alph{section}.\arabic{equation}}

\section*{Appendix}

\section{Normal modes of vibration}
\subsection{Force matrix and normal coordinates}
\label{app-force}
 
We consider a molecule with $N$ atoms, which are approximated by point
masses whose displacements from equilibrium due to vibrational motion
are encoded in $3N$ mass-weighted coordinates $q_i(t), i=1,..,N$,
where~$t$ is time.  We drop the explicit time dependence in what
follows, and use the notation $\dot {q_i}={\rm d}q_i/{\rm d}t$ and
$\ddot{q_i}={\rm d}^2q_i/{\rm d}t^2$. We also use Einstein's summation
convention, for instance, $\sum_{j=1}^{3N} A_{ij}\,v_j \equiv
A_{ij}\,v_j$. The $3N$ classical equations of motion are
\begin{equation} \label{ode}
\ddot{q_i}+F_{ij}q_j=0,\qquad i=1,..,3N\,,
\end{equation}
where the kinetic energy is given by $T=\frac{1}{2}\dot{q_i}^2$ and
the potential energy, Taylor expanded about its equilibrium position
and normalised so that it vanishes at equilibrium, boils down, in the
harmonic approximation, to the following expression
\begin{equation}
V=F_{ij}q_iq_j,
\end{equation}
with the $3N \times 3N$ force matrix or Hessian given by,
\begin{equation}
\left. F_{ij}=\frac{\partial^2 V}{\partial q_i \partial
  q_j}\right|_{q_k=q_k^0}, \qquad {\rm where}\,\,q_k^0\,\,\text{are
  the atoms' equilibrium positions.} 
\end{equation}
One is interested in changing basis from the general coordinates
$\{q_i\}$ to a new set of coordinates $\{ q'_k\}$ so that the set of
differential equations \eqref{ode} is equivalent to the set
\begin{equation}\label{normal}
\ddot{q}^{\,'}_k+\lambda_{(k)}q^{'}_k =0,\qquad k=1,..,3N\,,
\end{equation}
for some real values $\lambda_{(k)}$ to be determined. If the
orthogonal matrix of change of basis is~$\mu$,
i.e.~$q'_k=\mu_{ki}\,q_i$, one must have, after constructing $3N$
linear combinations of \eqref{ode} with coefficients $C_{ki}$,
\begin{equation}
C_{ki}\ddot{q_i}+C_{ki}F_{ij}q_j=0=\ddot{q}^{\,'}_k+\lambda_{(k)}q^{'}_k 
\end{equation}
where $C_{ki}=\mu_{ki}$ and $C_{ki}F_{ij}=\lambda_{(k)}\mu_{kj}$. This
implies that one must find $3N$ sets of $3N$ coefficients $C_{kj}$
such that
\begin{equation}\label{eigenvalue}
(F_{ij} -\lambda_{(k)}\delta_{ij})\,C_{ki}=0.
\end{equation}
It is clear from \eqref{eigenvalue} that, unless all coefficients are
zero, the $\lambda_{(k)}$ are eigenvalues of the force matrix. For
each such eigenvalue (fixed $k$), the equation~\eqref{eigenvalue}
allows to determine $3N$ coefficients $C_{ki}$. The $3N$ normal
coordinates $q^{\,'}_k=C_{ki}q_i$ are solutions of the differential
equations \eqref{normal}, so they read,
\begin{equation}
q^{\,'}_k=A_k\,\cos(\lambda_{(k)}^{1/2}\,t + \epsilon_{(k)}), \qquad k=1,..,3N\,,
\end{equation}
where $\epsilon_{(k)}$ are phases and $A_k$ are constants. Reverting
to the original system of coordinates,
\begin{equation}
q_i=C_{ij}\,q^{\,'}_j=C_{ij}\,A_j\,\cos(\lambda_{(j)}^{1/2}\,t + \epsilon_{(j)}), \qquad i=1,..,3N\,,
\end{equation}
and choosing all but one constant $A_k$ to be zero yields the $i^{th}$
normal mode of vibration
\begin{equation}
q_i=C_{ik}\,A_k\,\cos(\lambda_{(k)}^{1/2}\,t +
\epsilon_{(k)}),\qquad{\rm no\,summation\,on\,}k
\end{equation}
of frequency $\nu_{(k)}=\lambda^{1/2}_{(k)}/{2\pi}$.


\subsection{Group theory patterns of normal modes of vibration}
\label{app-patterns}

In this appendix we will briefly recall the method of decomposing 
vibration modes of a $G$-invariant $N$-atom molecule into irreducible
representations of the symmetry group $G$ (for more details 
see e.g.~\cite{Cornwell1}).  The goal is to find a new
basis~$\vec{q}^{\,\,'}$ for the atoms' positions, such that the action
of $G$ on~$\vec{q}^{\,\,'}$ decomposes into a sum of irreducible
representations. The action of the group $G$ on~$\vec{q}^{\,\,'}$ is called
the \emph{displacement representation}, and is simply given by the
tensor product
\begin{equation}
\label{e:disprep}
\Gamma^{\text{disp}}(g) = P(g) \otimes R(g)\,,\qquad g \in G\,.
\end{equation}
Equivalently, in terms of the components, we have $
(\Gamma^{\text{disp}})_{3m - 3 + i, 3 n - 3 + j} = P(g)_{mn} \otimes
R^{ij}(g)$, using the notation introduced above~\eqref{e:EOM}.  The
displacement representation matrices do not depend on the positions of
the masses, but only on the number of atoms. 

We denote with~$U$ the
matrix which achieves the coordinate transformation~$\vec{q}^{\,\,'} =
U \vec{q}$ such that it block-diagonalises the displacement
generators,
\begin{equation}
\label{e:decompGamma}
\Gamma'^{\,\text{disp}}(g) := U^{-1} \Gamma^{\text{disp}}(g) U = \oplus_{p} n_p \Gamma^p(g)\,.
\end{equation}
Here~$\Gamma^p(g)$ are the generators of the symmetry group $G$ in the
$p$-th representation. For every symmetry generator of the molecule
which leaves the potential invariant, we necessarily have
\begin{equation}\label{Fcommute}
\Gamma^{\text{disp}}(g)\, F = F\, \Gamma^{\text{disp}}(g),
\end{equation}
where $F$ is the force matrix introduced in appendix~\ref{app-force}.
By Schur's lemma, the block diagonal form of the right-hand side
of~\eqref{e:decompGamma},
\begin{equation}
\Gamma'^{\,\text{disp}}(g) = 
\begin{pmatrix}
\underbrace{\begin{matrix}
\Gamma^{p_1} &  & \\
 & \ddots & \\
 &  & \Gamma^{p_1}
\end{matrix}}_{n_{p_1}\times n_{p_1}} & & \\
& \underbrace{\begin{matrix}
\Gamma^{p_2} &  & \\
 & \ddots & \\
 &  & \Gamma^{p_2}
\end{matrix}}_{n_{p_2}\times n_{p_2}} & \\
& & \qquad\ddots\qquad
\end{pmatrix} 
\end{equation}
(in which all~$\Gamma^{p}$ are $d_p\times d_p$ matrices) thus implies
a similar block-diagonal form of the {$U$-transformed} force matrix~$F'
= U^{-1} F U$,
\begin{equation}
F' = \begin{pmatrix}
\underbrace{\begin{pmatrix}
F^{p_1}_{11} & \cdots  & F^{p_1}_{1\,n_{p_1}}\\
\vdots & \ddots & \vdots\\
F^{p_1}_{n_{p_1}\,1} & \cdots & F^{p_1}_{n_{p_1}\,n_{p_1}}
\end{pmatrix}}_{n_{p_1}\times n_{p_1}}\otimes
\mathbb{1}_{d_{p_1}\times d_{p_1}} & & \\
& \underbrace{\begin{pmatrix}
F^{p_2}_{11} & \cdots  & F^{p_2}_{1\,n_{p_2}}\\
\vdots & \ddots & \vdots\\
F^{p_2}_{n_{p_2}\,1} & \cdots & F^{p_2}_{n_{p_2}\,n_{p_2}}
\end{pmatrix}}_{n_{p_2}\times n_{p_2}}\otimes \mathbb{1}_{d_{p_2}\times d_{p_2}} & & \\
& & \qquad\ddots\qquad
\end{pmatrix} \,.
\end{equation}
Every irreducible representation~$\Gamma^{p}$ on the right-hand side
of~\eqref{e:decompGamma} thus corresponds to a set of~$n_p$ eigenvalues
of~$F$, each occurring with multiplicity~$d_p$.  

The coordinate transformation matrix~$U$ itself can be found using projection
operators. These are defined as
\begin{equation}
\big({\cal P}^p_{mn}\big)_{ab} 
  = \frac{d_p}{{\rm dim}\,G}\sum_{g\in G} \big(\Gamma^p(g)\big)^*_{nm}\,
  \big(\Gamma^{\text{disp}}(g)\big)_{ab}\,,\qquad
\text{$m,n=1,\ldots,d_p$}\,,\quad
\text{$a,b=1,\ldots,3N$}\,.
\end{equation}
(In the case of the icosahedral group the representations are all real
and the complex conjugation symbol can be ignored).  A crucial
property of these operators is that they provide us with basis vectors
transforming according to the representation~$\Gamma^p$. To see this,
let us transform a \emph{row} of the matrix~${\cal P}^p_{mn}$ by
acting on it with a group element~$h$ in the displacement
representation,
\begin{equation}
\begin{aligned}
\big({\cal P}^{p}_{m n}\big)'_{b a} 
  = \big(\Gamma^{\text{disp}}(h)\big)_{a c} \big({\cal P}^{p}_{mn}\big)_{b c}
  &=
\frac{d_p}{{\rm dim}\,G}\sum_{g} \big(\Gamma^p(g)\big)^*_{nm}\,
\big(\Gamma^{\text{disp}}(g)\,\Gamma^{\text{disp}}(h^{-1})\big)_{b a}\\[1ex]
&=
    \frac{d_p}{{\rm dim}\,G}\sum_{g'=g h^{-1}} \big(\Gamma^p(g')\big)^*_{nk}\,
                                   \big(\Gamma^p(h)\big)^*_{km}\,
    \big(\Gamma^{\text{disp}}(g')\big)_{b a}\\[1ex]
&= \big(\Gamma^p(h)\big)^*_{km}\, \big({\cal P}^{p}_{kn}\big)_{b a}\,.
\end{aligned}
\end{equation}
This shows that the rows of the matrices~${\cal P}^p_{mn}$ transform
into each other according to the $p$-th representation. In order to
construct a transformation matrix which takes us to the
symmetry-adapted basis, we thus have to construct the span of all row
vectors in all~${\cal P}^p_{mn}$ matrices (for fixed~$p$ and
all~$m,n$). In fact, the normalised row-vectors of the
matrices~${\cal P}^p_{mm}$, with $m=1\ldots d_p$, together already
span an~$n_p$-dimensional subspace transforming as~$\Gamma^p$. The
matrix~$U^T$ is now constructed by taking an orthonormal basis in each
of these subspaces, and then using these basis vectors to populate the
rows of~$U^T$.  The matrix~$U$ then also block-diagonalises the force
matrix~$F$. Note once more that~$U$ only depends on the number of
capsid proteins, not on their precise positions. For more details we
refer the reader to e.g.~\cite{Cornwell1}.

If one is only interested in the multiplicities of the various
irreducible representations, the~$U$ matrix is not needed and it is
more convenient to use the character formula
\begin{equation}
n_p = \frac{1}{\dim G} \sum_{g\in G} \chi^{*\,\text{disp}}(g)\,
\chi^p(g)\,.
\end{equation}
Here the character of a group element~$g$ in a given representation
is obtained by computing the trace of the associated matrix,
$\chi(g) = \Tr\big( R(g)\big)$. For the displacement
representation~\eqref{e:disprep} the character becomes
\begin{equation}
\chi^{\text{disp}}(g) = \Tr\big( P(g) \big)\cdot \Tr\big( R(g)\big)
= \pm (\text{\# of proteins unmoved by~$g$})\cdot (1+2\cos\theta)\,,
\end{equation}
where~$\theta$ is the angle of the rotation associated with~$g$, and
the minus sign is taken whenever~$g$ involves an inversion. The
characters of irreducible representations of finite groups are widely
available in the literature, and are reproduced in
appendix~\ref{app-characters} for the icosahedral group~$H_3$.  This
fixes the multiplicities~$n_p$ and can be used as a check on the
intermediate step of the calculation of the projection operators.

\begin{figure}[t]
\begin{center}
\includegraphics[width=.25\textwidth]{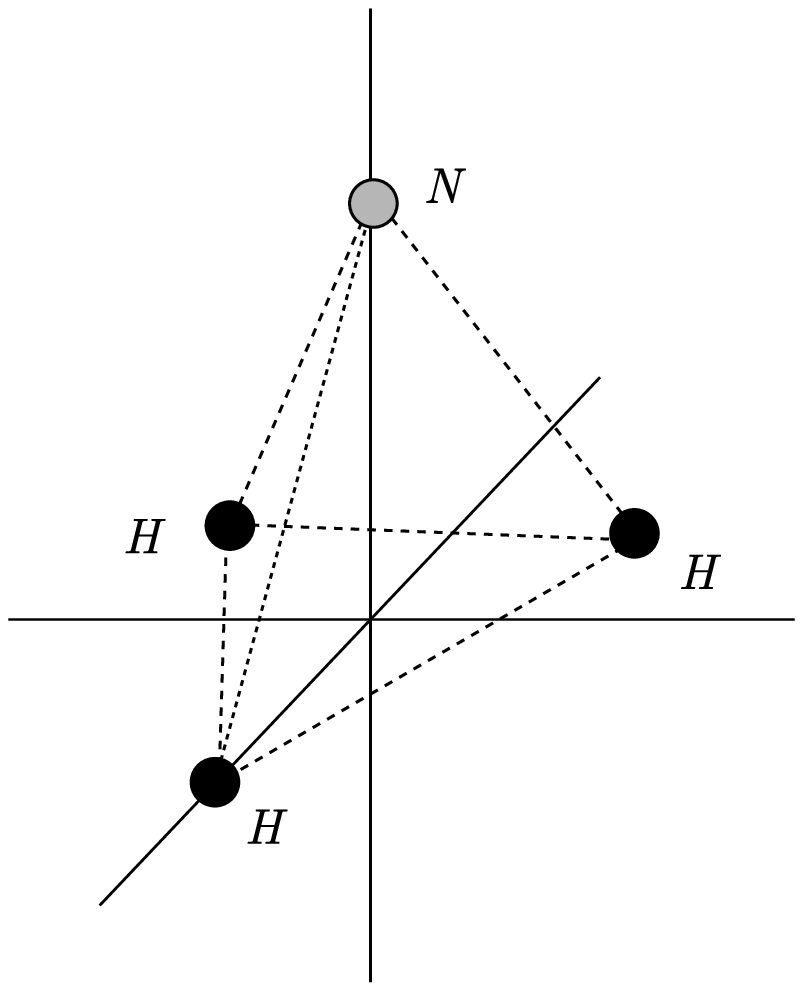}
\end{center}
\vspace{-2ex}
\caption{The structure of ammonia.\label{f:ammonia}}
\end{figure}

Let us, to conclude, illustrate the technique described above for the case
of the ammonia molecule (see \figref{f:ammonia}), whose point group is~$C_{3v}$ and whose
displacement representation matrices can be shown to satisfy
\begin{equation} 
\label{normalmodes}
\Gamma^{\text{disp}\,'}(g)=U^{-1}\,\Gamma^{\text{disp}}(g)\,U=3\,\Gamma^1(g)+\Gamma^{1'}(g)+4\Gamma^2(g),\quad \forall g \in C_{3v}.
\end{equation}
Hence the multiplicities are~$n_1=3$, $n_2=1$ and $n_3=4$.
The matrix $F^{\,'}$ can be partitioned in nine blocks of dimensions
$n_i \times n_j, i,j=1,2,3$ to match the structure of the displacement
representation~\eqref{normalmodes}, and use the following notations
 \begin{equation} \label{matrixF}
F^{\,'}= 
\left( \begin{array}{cccc|ccc|ccccc}
F^{\,'11}_{11}&F^{\,'11}_{12}&F^{\,'11}_{13}&&&F^{\,'12}_{11}&&&F^{\,'13}_{11}&F^{\,'13}_{12}&F^{\,'13}_{13}&F^{\,'13}_{14}\\
F^{\,'11}_{21}&F^{\,'11}_{22}&F^{\,'11}_{23}&&&F^{\,'12}_{21}&&&F^{\,'13}_{21}&F^{\,'13}_{22}&F^{\,'13}_{23}&F^{\,'13}_{24}\\
F^{\,'11}_{31}&F^{\,'11}_{32}&F^{\,'11}_{33}&&&F^{\,'12}_{31}&&&F^{\,'13}_{31}&F^{\,'13}_{32}&F^{\,'13}_{33}&F^{\,'13}_{34}\\
&&&&&&&&&&&\\
\hline
&&&&&&&&&&&\\
F^{\,'21}_{11}&F^{\,'21}_{12}&F^{\,'21}_{13}&&&F^{\,'22}_{11}&&&F^{\,'23}_{11}&F^{\,'23}_{12}&F^{\,'23}_{13}&F^{\,'23}_{14}\\
&&&&&&&&&&&\\
\hline
&&&&&&&&&&&\\
F^{\,'31}_{11}&F^{\,'31}_{12}&F^{\,'31}_{13}&&&F^{\,'32}_{11}&&&F^{\,'33}_{11}&F^{\,'33}_{12}&F^{\,'33}_{13}&F^{\,'33}_{14}\\
F^{\,'31}_{21}&F^{\,'31}_{22}&F^{\,'31}_{23}&&&F^{\,'32}_{21}&&&F^{\,'33}_{21}&F^{\,'33}_{22}&F^{\,'33}_{23}&F^{\,'33}_{24}\\
F^{\,'31}_{31}&F^{\,'31}_{32}&F^{\,'31}_{33}&&&F^{\,'32}_{31}&&&F^{\,'33}_{31}&F^{\,'33}_{32}&F^{\,'33}_{33}&F^{\,'33}_{34}\\
F^{\,'31}_{41}&F^{\,'31}_{42}&F^{\,'31}_{43}&&&F^{\,'32}_{41}&&&F^{\,'33}_{41}&F^{\,'33}_{42}&F^{\,'33}_{43}&F^{\,'33}_{44}\\
\end{array}\right) \nonumber
\end{equation}
In the above, the expressions $F^{\,'pq}_{\alpha\beta},
\alpha=1,..,n_p; \beta=1,..,n_q; p,q=1,2,3$ are matrices of dimension
$d_p \times d_q$.

The equality $\Gamma^{\text{disp} \,'}(g)
F{\,'}=F^{\,'}\Gamma^{\text{disp} \,'}(g)\quad \forall g \in C_{3v}$
is equivalent to
\begin{equation}
\Gamma^p(g)F^{\,'pq}_{\alpha\beta}=F^{\,'pq}_{\alpha\beta}\Gamma^q(g),\quad \forall g \in C_{3v},
\end{equation}
for $ \alpha=1,..,n_p; \beta=1,..,n_q; p,q=1,2,3$. We are now in a position to make good use of Schur's Lemma. If $p \neq q$, the irreducible representations $\Gamma^p$ and $\Gamma^q$ are not equivalent, and this implies
\begin{equation}
F^{\,'pq}_{\alpha\beta}=\left\{ \begin{array}{c} 0,\qquad {\rm if}\,\,p\neq q\\
                                                             f_{\alpha\beta}^p\,{\bf I}_{d_p},\,\, {\rm if}\,\,
                                                             p=q\end{array}\right.
                                                             \end{equation}
where $f_{\alpha\beta}^p$ are a set of $(n_p)^2$ complex numbers obeying $f_{\alpha \beta}^p=(f_{\beta \alpha}^p)^*$ as $F$, and hence $F^{\,'}$  is hermitian. So Schur's Lemma and the hermiticity of $F^{\,'}$ allow us to write
 \begin{equation} \label{matrixschur}
F^{\,'}=\left( \begin{array}{cccc|ccc|cccccccc}
f^1_{11}&f^1_{12}&f^1_{13}&&&&&&&&&&&&\\
(f^1_{12})^*&f^1_{22}&f^1_{23}&&&&&&&&&&&&\\
(f^1_{13})^*&(f^1_{23})^*&f^1_{33}&&&&&&&&&&&&\\
&&&&&&&&&&&&&&\\ \hline
&&&&&&&&&&&&&&\\
&&&&&f^2_{11}&&&&&&&&&\\
&&&&&&&&&&&&&&\\ \hline
&&&&&&&&&&&&&&\\
&&&&&&&f^3_{11}&0&f^3_{12}&0&f^3_{13}&0&f^3_{14}&0\\
&&&&&&&0&f^3_{11}&0&f^3_{12}&0&f^3_{13}&0&f^3_{14}\\
&&&&&&&(f^3_{12})^*&0&f^3_{22}&0&f^3_{23}&0&f^3_{24}&0\\
&&&&&&&0&(f^3_{12})^*&0&f^3_{22}&0&f^3_{23}&0&f^3_{24}\\
&&&&&&&(f^3_{13})^*&0&(f^3_{23})^*&0&f^3_{33}&0&f^3_{34}&0\\
&&&&&&&0&(f^3_{13})^*&0&(f^3_{23})^*&0&f^3_{33}&0&f^3_{34}\\
&&&&&&&(f^3_{14})^*&0&(f^3_{24})^*&0&(f^3_{34})^*&0&f^3_{44}&0\\
&&&&&&&0&(f^3_{14})^*&0&(f^3_{24})^*&0&(f^3_{34})^*&0&f^3_{44}\\
\end{array}\right)\nonumber
\end{equation}
where all entries not explicitly written are zero.  After a further
similarity transformation that reshuffles rows 7, 9 and 11, and then
columns 7, 9 and 11, we arrive at (noticing also that $F$ and $U$ are
real)
 \begin{equation} \label{matrixschur2}
F^{\,''}=\left( \begin{array}{cccc|ccc|cccc|cccc}
f^1_{11}&f^1_{12}&f^1_{13}&&&&&&&&&&&&\\
f^1_{12}&f^1_{22}&f^1_{23}&&&&&&&&&&&&\\
f^1_{13}&f^1_{23}&f^1_{33}&&&&&&&&&&&&\\
&&&&&&&&&&&&&&\\ \hline
&&&&&&&&&&&&&&\\
&&&&&f^2_{11}&&&&&&&&&\\
&&&&&&&&&&&&&&\\ \hline
&&&&&&&&&&&&&&\\
&&&&&&&f^3_{11}&f^3_{12}&f^3_{13}&f^3_{14}&&&&\\
&&&&&&&f^3_{12}&f^3_{22}&f^3_{23}&f^3_{24}&&&&\\
&&&&&&&f^3_{13}&f^3_{23}&f^3_{33}&f^3_{34}&&&&\\
&&&&&&&f^3_{14}&f^3_{24}&f^3_{34}&f^3_{44}&&&&\\
&&&&&&&&&&&&&&\\ \hline
&&&&&&&&&&&&&&\\
&&&&&&&&&&&f^3_{11}&f^3_{12}&f^3_{13}&f^3_{14}\\
&&&&&&&&&&&f^3_{12}&f^3_{22}&f^3_{23}&f^3_{24}\\
&&&&&&&&&&&f^3_{13}&f^3_{23}&f^3_{33}&f^3_{34}\\
&&&&&&&&&&&f^3_{14}&f^3_{24}&f^3_{34}&f^3_{44}\\
\end{array}\right)
\end{equation}
So we have managed to transform the force matrix $F$ into an
equivalent matrix, $F^{\,''}$, which is block diagonal. It now remains
to calculate the eigenvalues and the eigenvectors of the smaller size
matrices appearing in the block diagonal form to obtain the normal
modes of vibration. Note that each eigenvalue of $F$ is associated
with one irreducible representation $\Gamma^p$ of $C_{3v}$, and each
such eigenvalue is $d_p$-fold degenerate, where $d_p$ is the dimension
of $\Gamma^p$.


\section{The Tirion and RTB approximations}
\label{app-Tirion}

In order to understand the approximations made with the Tirion
potential and the rotations-translations of blocks (RTB) method, let us
consider a simple example. This example models two protein chains,
each individually bound together by strong covalent bonds, and
mutually interacting through weak Van der Waals bonds. This is a
caricature of the real world situation, but useful to understand the
logic of these two approximations.

\begin{figure}[ht]
\begin{center}
\includegraphics[width=.5\textwidth]{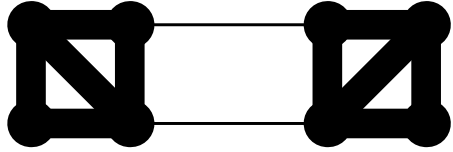}
\end{center}
\vspace{-5ex}
\caption{Two simplified protein chains whose mutual interaction is very weak compared to the interatomic forces within each protein. The thickness of the lines symbolises the strength of the bonds.\label{simple-model}}
\end{figure}

The model we will consider lives in two dimensions. We have two
protein chains of 4 atoms each, positioned on the nodes of a
square. These two chains interact weakly, as in \figref{simple-model}.

\begin{figure}[t]
\begin{center}
\vspace{2ex}
\includegraphics[width=.5\textwidth]{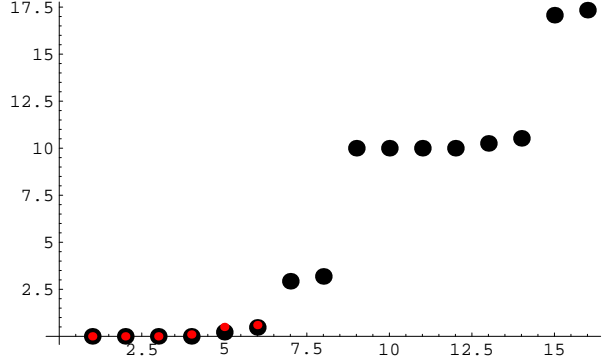}
\vspace{-2ex}
\end{center}
\caption{Exact frequency spectrum of the two protein chains' system. \label{spectrum-example}}
\end{figure}

This ``protein'' has one zero-mode in addition to the three trivial
ones (translation and rotation). This additional zero-mode corresponds
to shearing motion of the central square. In order to exhibit the
consequences of a large hierarchy of interaction strengths, we will
take the ratio of the strong to weak bonds to be~10. The force matrix
of this model is computed as described earlier, keeping track of the
different spring constants.  The exact spectrum of this model is given
by the black dots in the plot of \figref{spectrum-example}.

In the ``rotations-translations of blocks'' (RTB) method, one computes
the spectrum of the low-frequency modes by grouping together clusters
(blocks) of atoms which form relatively rigid structures. By making
these clusters completely rigid and keeping only their translation and
rotation degrees of freedom, the total computational complexity is
reduced. In our example above, we will use the two protein chains as
blocks. There are two translations for each block, and one rotation,
yielding a total of 6 coordinates. These are given by
\begin{equation}
\begin{aligned}
t_x^1 &= \frac{1}{2}(e_x^1 + e_x^2 + e_x^7 + e_x^8)\,,\\[1ex]
t_y^1 &= \frac{1}{2}(e_y^1 + e_y^2 + e_y^7 + e_y^8)\,,\\[1ex]
t_x^2 &= \frac{1}{2}(e_x^3 + e_x^4 + e_x^5 + e_x^6)\,,\\[1ex]
t_y^2 &= \frac{1}{2}(e_y^3 + e_y^4 + e_y^5 + e_y^6)\,,
\end{aligned}\qquad
\begin{aligned}
r^1   &= \frac{1}{2}(e_x^1 + e_x^2 - e_x^7 - e_x^8)\,,\\[1ex]
r^2   &= \frac{1}{2}(e_x^3 + e_x^4 - e_x^5 - e_x^6)\,.
\end{aligned}
\end{equation}
Note that the rotation coordinates are so simple because we only look
at the linearised approximation.  The resulting projection matrix~$P$
reads

\begin{equation}
P = \frac{1}{2}
\left(\begin{array}{rrrrrrrrrrrrrrrr}
1 & 0 & 1 & 0 & 0 & 0 & 0 & 0 & 0 & 0 & 0 & 0 & 1 & 0 & 1 & 0\\
0 & 1 & 0 & 1 & 0 & 0 & 0 & 0 & 0 & 0 & 0 & 0 & 0 & 1 & 0 & 1\\
0 & 0 & 0 & 0 & 1 & 0 & 1 & 0 & 1 & 0 & 1 & 0 & 0 & 0 & 0 & 0\\
0 & 0 & 0 & 0 & 0 & 1 & 0 & 1 & 0 & 1 & 0 & 1 & 0 & 0 & 0 & 0\\
1 & 0 & 1 & 0 & 0 & 0 & 0 & 0 & 0 & 0 & 0 & 0 & -1 & 0 & -1 & 0\\
0 & 0 & 0 & 0 & 1 & 0 & 1 & 0 & -1 & 0 & -1 & 0 & 0 & 0 & 0 & 0
\end{array}\right)
\end{equation}

\noindent This projection matrix is orthogonal, i.e.~it satisfies
\begin{equation}
P \cdot P^T = 1_{6\times 6}.
\end{equation}
The force matrix for the reduced system now reads
\begin{equation}
P \cdot F \cdot P^T = 
 \frac{1}{4}
 \begin{pmatrix}
 1 & 0 & -1 & 0 & 0 & 0\\
 0 & 0 & 0 & 0 & 0 & 0\\
 -1 & 0 & 1 & 0 & 0 & 0\\
 0 & 0 & 0 & 0 & 0 & 0\\
 0 & 0 & 0 & 0 & 11 & -1 \\
 0 & 0 & 0 & 0 & -1 & 11
 \end{pmatrix}\,.
\end{equation}
The eigenvalues of this matrix are
\begin{equation}
0, \,\,0, \,\,0,\, \,\frac{1}{2},\,\, \frac{5}{2},\,\, \frac{6}{2} \,.
\end{equation}
These frequencies are, as expected~\cite{Sanejouand:2007a}, larger
than the actual frequencies. They are depicted by the small red dots
in the figure above, where a scaling by a factor of~$1/5$ was applied.
We thus see that this method does a decent job at reproducing the
low-frequency spectrum of the exact model, and correctly throws away
all the high-frequency modes.

\begin{figure}[ht]
\begin{center}
\includegraphics[width=.3\textwidth]{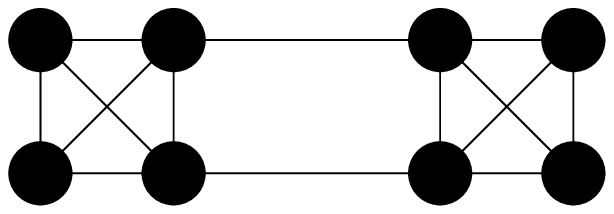}
\includegraphics[width=.3\textwidth]{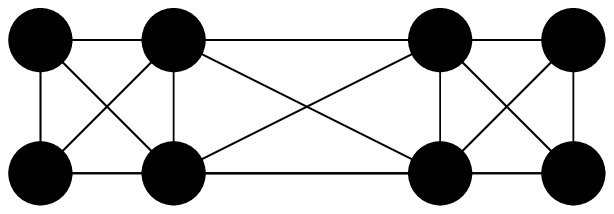}
\includegraphics[width=.3\textwidth]{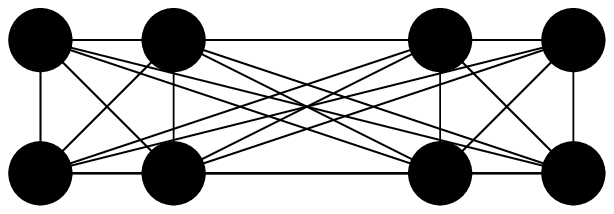}
\end{center}
\vspace{-2ex}
\caption{Visualisation of the Tirion bonds of the toy model protein for three
  cutoff radii.\label{Tirioncutoffs}.}
\end{figure}

\begin{figure}[ht]
\vspace{3ex}
\begin{center}
\includegraphics[width=.3\textwidth]{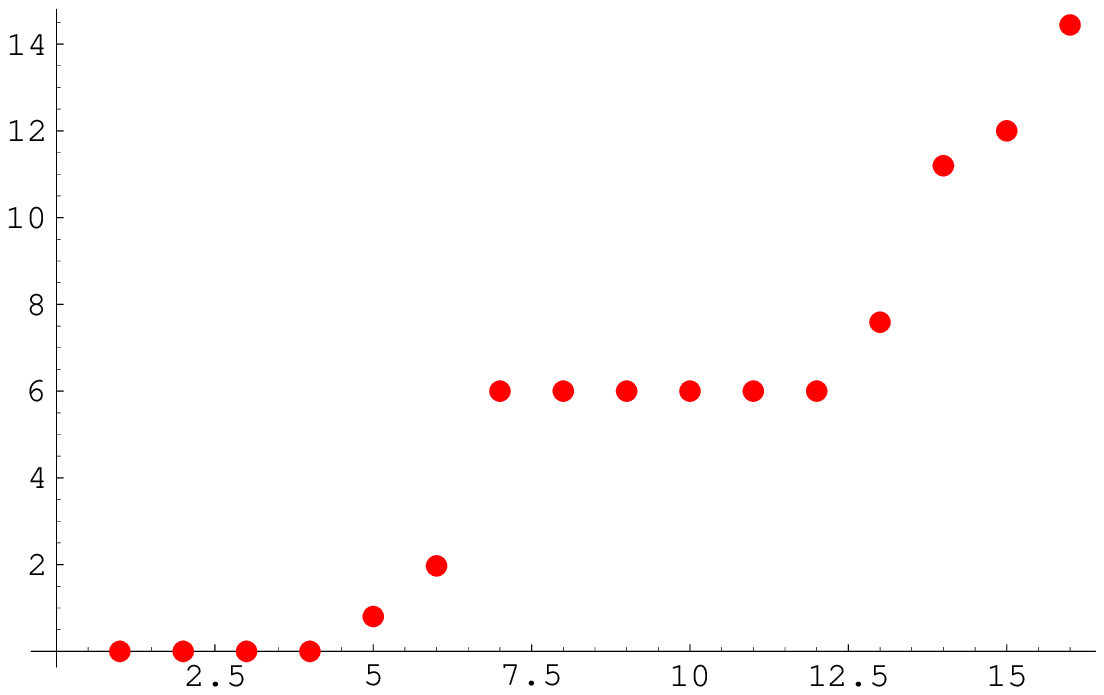}
\includegraphics[width=.3\textwidth]{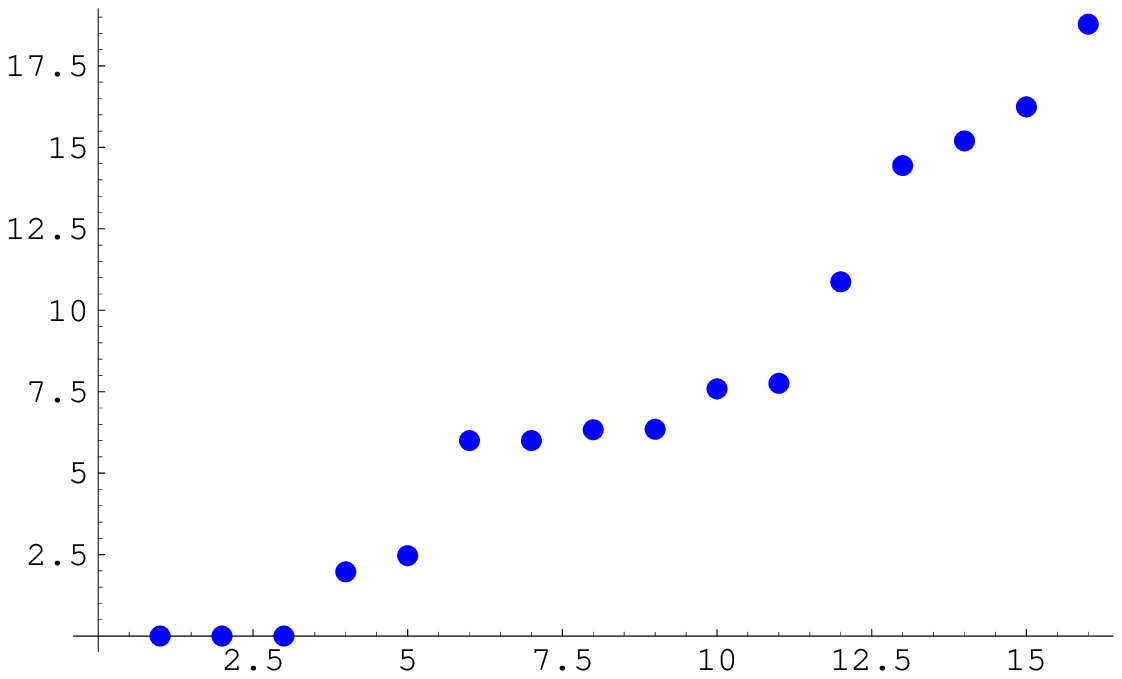}
\includegraphics[width=.3\textwidth]{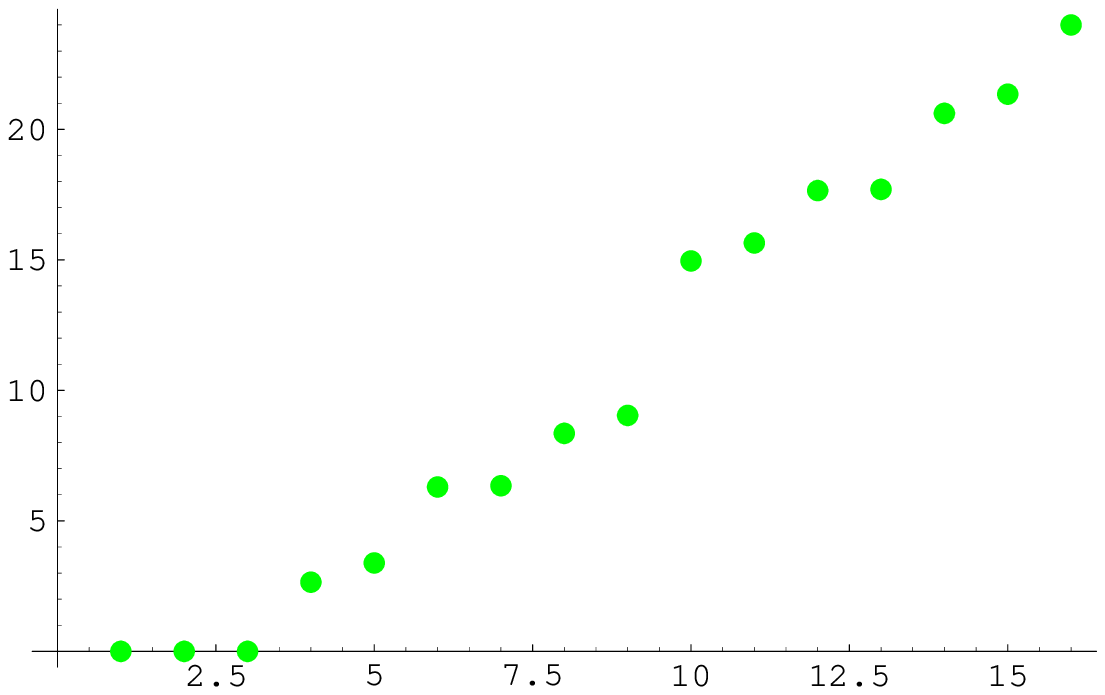}
\end{center}
\caption{The exact frequency spectra for the Tirion potential with three different cutoffs, for the two protein chains' system.\label{spectra-Tirion}}
\end{figure}

Now consider the Tirion approximation. In this approximation, we
forget about all details of the bonds, and instead write down a
potential which depends on only one overall spring constant (to be
fixed by hand). Moreover, the potential depends on a cutoff
radius. In \figref{Tirioncutoffs} we show three configurations for three cutoff
radii~$R_c=2.1$, $3.1$ and $5.1$. 
We arbitrarily fix the spring constant to be~$3$, so the plots of the
spectrum should be read modulo an overall normalisation.
For the three
cutoff radii mentioned above, the spectra are presented in \figref{spectra-Tirion}.
From these plots it is obvious that this method can produce results
which are both quantitatively and qualitatively wrong. In particular,
it fails to reproduce the separation between high-frequency and
low-frequency modes; instead, one is ``supposed to know'' where to
stop trusting the Tirion spectrum. Note, however, that a sufficiently
small cutoff can \emph{still} change the zero-mode structure and
introduce instabilities (the standard solution, see
e.g.~\cite{Atilgan:2001a}, is to simply increase the cutoff length
until the capsid is stable, but this seems rather arbitrary).

\section{Irreducible representations and characters of the icosahedral group}
\label{app-characters} 
 
 For completeness we list here the irreducible representations of the
 icosahedral group, based on those of~\cite{Hoyle:2004a}.  There are
 ten irreducible representations, with generators labelled as~$M^{i
   \pm}_{g}$, where~$i=1\ldots 5$ (labelling the representation) and
 $g\in\{ -, 2, 5\}$ (indicating the inversion element, the two-fold
 and the five-fold generator respectively). They thus come in pairs,
 in which the action of the inversion element in the `$+$'
 representation is trivial, while it inverts all coordinates in the
 `$-$' representation. That is,
\begin{equation}
M^{i +}_{-} = 1\,,\qquad
M^{i -}_{-} = -1\,.
\end{equation}
The form of the two-fold and the five-fold rotation generators is
listed below. Here~$\tau = \tfrac{1}{2}(1+\sqrt{5})$ denotes the
golden ratio.
\begin{subequations}
\begin{equation}
\begin{aligned}
M^{1\pm}_2 &= 1\,,\qquad &
M^{1\pm}_5 &= 1\,,\\[1ex]
M^{2\pm}_2 &= \begin{pmatrix}
-1 & 0 & 0 \\
0 & 1 & 0 \\
0 & 0 & -1
\end{pmatrix}
\,,\qquad &
M^{2\pm}_5 &= \frac{1}{2}\begin{pmatrix}
\tau^{-1} & -\tau & 1 \\
\tau & 1 & \tau^{-1} \\
-1 & \tau^{-1} & \tau
\end{pmatrix}\,,
\end{aligned}
\end{equation}
\begin{equation}
\begin{aligned}
M^{3\pm}_{2} &= \begin{pmatrix}
-1 & 0 & 0 \\
 0 & 1 & 0 \\
 0 & 0 & -1
\end{pmatrix}\,,\qquad &
M^{3\pm}_{5} &= \frac{1}{2}\begin{pmatrix}
-\tau & -\tau^{-1} & 1\\
\tau^{-1} & 1 & \tau\\
-1 & \tau & -\tau^{-1}
\end{pmatrix}\,,\\[1ex]
M^{4\pm}_{2} &= \begin{pmatrix}
0 & 0 & 1 & 0 \\
0 & 0 & 0 & 1 \\
1 & 0 & 0 & 0 \\
0 & 1 & 0 & 0
\end{pmatrix}\,,\qquad &
M^{4\pm}_{5} &= \begin{pmatrix}
-1 & 1 & 0 & 0 \\
-1 & 0 & 1 & 0 \\
-1 & 0 & 0 & 1 \\
-1 & 0 & 0 & 0
\end{pmatrix}\,,\\[1ex]
M^{5\pm}_{2} &= \begin{pmatrix}
0 & -1 & 1 & 0 & 0 \\
0 & -1 & 0 & 0 & 0 \\
1 & -1 & 0 & 0 & 0 \\
0 & -1 & 0 & 1 & 0 \\
0 & -1 & 0 & 0 & 1 
\end{pmatrix}\,,\qquad &
M^{5\pm}_{5} &= \begin{pmatrix}
1 & 0 & 0 & 0 & 1 \\
0 & 0 & 0 & 0 & -1\\
0 & 0 & 0 & 1 & -1\\
0 & 1 & 0 & 0 & -1\\
0 & 0 & 1 & 0 & -1
\end{pmatrix}\,.\\[1ex]
\end{aligned}
\end{equation}
\end{subequations}
For the decomposition of the displacement representation into
irreducible representations we frequently make use of the characters
of the icosahedral group. These are listed in
table~\ref{character-table}, both for $H_3$ as well as its proper
rotational subgroup~${\cal I}$.
\begin{table}[ht]
\vspace{2ex}
\begin{center}
\newcolumntype{K}{>{\raggedleft\arraybackslash}X}
\begin{tabularx}{\textwidth}{c*{10}{K}}
conj.~class              & ${\cal C}(e)$ & ${\cal C}(g_5)$ & ${\cal C}(g_5^2)$ &
  ${\cal C}(g_3)$ & ${\cal C}(g_2)$ 
& ${\cal C}(g_0)$ & ${\cal C}(g_0 g_5)$ & ${\cal C}(g_0 g_5^2)$ &
${\cal C}(g_0 g_3)$ & ${\cal C}(g_0 g_2)$ \\[.2ex]
size & 1 & 12 & 12 & 20 & 15 & 1 & 12 & 12 & 20 & 15 \\[1ex]
\hline \\[-.5ex]
$\Gamma^1_{+}$  & 1 & 1       & 1       & 1    & 1 & 1 & 1 & 1 & 1 & 1 \\
$\Gamma^2_{+}$  & 3 & $\tau$  & $\tau'$ & 0    & $-1$ & 3 & $\tau$ & $\tau'$ & 0 & $-1$\\
$\Gamma^3_{+}$  & 3 & $\tau'$ & $\tau$  & 0    & $-1$ & 3 & $\tau'$ & $\tau$ & 0 & $-1$ \\
$\Gamma^4_{+}$  & 4 & $-1$    & $-1$    & 1    & 0 & 4 & $-1$ & $-1$ & 1 & 0 \\
$\Gamma^5_{+}$  & 5 & 0       & 0       & $-1$ & 1 & 5 & 0 & 0 & $-1$ & 1 \\[1ex]
\cline{1-6}\\[-1.4ex]
$\Gamma^1_{-}$ & 1 & 1 & 1 & 1 & 1 & $-1$ & $-1$ & $-1$ & $-1$ & $-1$\\
$\Gamma^3_{-}$ & 3 & $\tau$ & $\tau'$ & 0 & $-1$ & $-3$ & $-\tau$ & $-\tau'$ & $0$ & $1$\\
$\Gamma^{3'}_{-}$ & 3 & $\tau'$ & $\tau$ & 0 & $-1$ & $-3$ & $-\tau'$ & $-\tau$ & $0$ & $1$\\
$\Gamma^4_{-}$ & 4 & $-1$ & $-1$ & 1 & 0 & $-4$ & $1$ & $1$ & $-1$ & $0$\\
$\Gamma^5_{-}$ & 5 & 0 & 0 & $-1$ & 1 & $-5$ & 0 & 0 & 1 & $-1$
\end{tabularx}
\end{center}
\caption{Characters of the irreducible representations of the
  icosahedral group~$H_3$ and the subgroup ${\cal I}$ which does not
  include the inversion element (top left quadrant). The notation used
  is~$\tau= \frac{1}{2}(1+\sqrt{5})$ and
  $\tau'=\frac{1}{2}(1-\sqrt{5})$, and~${\cal C}(g)$ denotes all
  elements in the conjugacy class of the
  element~$g$.}\label{character-table}

\end{table}

\vfill\eject
\begin{small}

\begingroup\raggedright\endgroup

\end{small}

\end{document}